\documentclass{aa}  
\usepackage{tabularx}
\usepackage{graphicx}
\usepackage{txfonts}
\begin{document} 

 \title{Molecular gas kinematics in the nuclear region of nearby Seyfert galaxies with ALMA}

%\subtitle{Molecular gas in active galaxies}
\titlerunning{Molecular gas kinematics in Seyfert galaxies}

   \author{A. Bewketu Belete
          \inst{1}
          \and 
          P. Andreani\inst{2}
          \and
          J. A. Fernández-Ontiveros\inst{3}
          \and 
          E. Hatziminaoglou\inst{2}
          \and
          F. Combes\inst{4}
          \and
          M. Sirressi\inst{5}
          \and
          R. Slater\inst{6,7}
          \and
          C. Ricci\inst{8}
          \and
          K. Dasyra\inst{9,10}
          \and
          C. Cicone\inst{11}
          \and 
          S. Aalto\inst{12}
          \and
          L. Spinoglio \inst{3}
          \and
          M. Imanishi\inst{13}
           \and
          J. R. De Medeiros\inst{1}
          %\and
         }

   \institute{Departamento de F\'isica Te\'orica e Experimental, Universidade Federal do Rio Grande do Norte, Natal, RN 59078-970, Brazil\\
    \email{asnakew@fisica.ufrn.br}
           \and
             European Southern Observatory, Karl-Schwarzschild-Straße 2, D–85748, Garching, Germany
                      \and
                     Istituto di Astrofisica e Planetologia Spaziali (INAF–IAPS), Via Fosso del Cavaliere 100, I–00133 Roma, Italy
                     \and
                    Observatoire de Paris, LERMA, College de France, PSL University, CNRS, Sorbonne University, F-75014 Paris, France
                     \and
                    Department of Astronomy and Oskar Klein Centre; Stockholm University; AlbaNova 106 91 Stockholm, Sweden
                     \and
                     Departamento de Astronomía, Universidad de La Serena, Av. Juan Cisternas 1200, La Serena, Chile
                     \and
                     Direcci\'on de Formaci\'on General, Facultad de Educaci\'on y Cs. Sociales, Universidad Andres Bello, Sede Concepci\'on, %autopista Concepci\'on-Talcahuano 7100,
 Talcahuano, Chile 
                     \and
                     N\'ucleo de Astronom\'ia, Facultad de Ingenier\'ia y Ciencias Universidad Diego Portales (UDP) Santiago de Chile
                     \and
                     Dep. of Astrophysics, Astronomy \& Mechanics, Faculty of Physics, National and Kapodistrian University of Athens, Panepis-timiopolis, Zografou 15784, Greece
                     \and
                     National Observatory of Athens, Institute for Astronomy, Astrophysics, Space Applications and Remote Sensing, Penteli 15236Athens, Greece
                     \and
                     Institute of Theoretical Astrophysics, University of Oslo, PO Box 1029, Blindern 0315, Oslo, Norway
                     \and
                     Chalmers University of Technology, Department of Earth and Space Sciences, Onsala Space Observatory, 439 92 Onsala, Sweden
                     \and
                     National Astronomical Observatory of Japan 2-21-1 Osawa, Mitaka, Tokyo 181-8588, Japan
                     }
   \date{Received xxx; accepted yyy}

  \abstract
    {The study of the distribution, morphology and kinematics of cold molecular gas in the nuclear and circumnuclear regions of active galactic nuclei (AGN) helps to characterise and hence to quantify the impact of the AGN on the host galaxy over its lifetime.}
   {We present the analysis of the molecular gas in the nuclear regions of three Seyfert galaxies, NGC 4968, NGC 4845, and MCG-06-30-15, with the help of ALMA observations of the CO(2-1) emission line. The aim is to determine the kinematics of the gas in the central ($\sim$ 1 kpc) region, and thereby to probe AGN nuclear fueling and feedback.}
   {We use two different softwares, namely the 3D-Based Analysis of Rotating Object via Line Observations ($^{3D}$BAROLO) and DiskFit, to model the kinematics of the gas in the molecular disc, and thereby to determine the gas rotation and any kinematical perturbations.}
   {Circular motions dominate the kinematics of the molecular gas in the central discs, mainly in NGC 4845 and MCG-06-30-15, however there is a clear evidence of non-circular motions in the central ($\sim$ 1 kpc) region of NGC 4845 and NGC 4968. The strongest non-circular motion is detected in the inner disc of NGC 4968, mainly along the minor kinematic axis, with velocity $\sim 115\, \rm{km\,s^{-1}}$.
 Of all DiskFit models, the bisymmetric model is found to give the best-fit for NGC 4968 and NGC 4845, indicating that the observed non-circular motions in the inner disc of these galaxies could be due to the nuclear barred structure, the gas streaming in elliptical orbits aligned along the bar. 
If the dynamics of NGC 4968 is modeled as a corotation pattern just outside of the bar, the bar pattern speed turns out to be at $\Omega_b$ = $52\, \rm{km\,s^{-1}\,kpc^{-1}}$ the corotation is set at 3.5 kpc and the inner Lindblad resonance (ILR) ring at R = 300pc corresponding to the CO emission ring.
   In the NGC~4968 galaxy, the torques exerted on the gas by the bar are positive in the centre, within the gas nuclear ring, and negative outside.  This shows that the gas is transiently trapped in the inner Lindblad resonance.
The comparison of the CO intensity maps with the map of the cold dust emission shows the absence of CO in the centre of NGC 4968 and that the dust distribution and the CO emission in and around the centre of NGC 4845 have similar extension. The 1.2 mm ALMA continuum is peaked and compact in NGC 4968 and MCG-06-30-15, but their CO(2-1) has an extended distribution. Allowing the CO-to-H$_{2}$ conversion factor $\alpha_{CO}$ between 0.8 and 3.2, typical of nearby galaxies of the same type, the molecular mass M(H$_{2}$) is estimated to be $\sim 3-12\times 10^{7} ~{\rm M_\odot}$ (NGC 4968), $\sim 9-36\times 10^{7}~ {\rm M_\odot}$ (NGC 4845), and $\sim 1-4\times 10^{7}~ {\rm M_\odot}$ (MCG-06-30-15). }
    {We conclude that the observed non-circular motions in the molecular disc of NGC 4968 and likely that seen in NGC 4845 is due to the presence of the bar in the nuclear region. We discuss the possibility that the observed pattern in the kinematics be due to the presence of the AGN and this might be the case for NGC4845. At the current spectral and spatial resolution and sensitivity we cannot claim any strong evidence in these sources of the long sought feedback/feeding effect due to the AGN presence.}
%\end{abstract}     
   \keywords {galaxies: active; galaxies: individual: NGC 4968, NGC 4845, MCG-06-30-15; 
   galaxies: ISM; galaxies: kinematics and dynamics; galaxies: nuclei; galaxies: spiral
   }
\maketitle
%
%-------------------------------------------------------------------
\section{Introduction}
Powerful active galactic nuclei (hereafter, AGNs) can drive fast outflows which affect the properties and kinematics of the ambient gas in the nuclear and circumnuclear regions of their host galaxies. By heating the gas, or/and expelling it from the central regions, they may affect the star formation histories of the host galaxy bulge (see e.g., \citealt{fluetsch2019cold}; \citealt{oosterloo2017properties}; \citealt{2017A&A...601A.143F}; \citealt{morganti2017many} and references therein). Even though the mechanisms that can produce molecular outflows are still uncertain (\citealt{morganti2017many}), cold molecular gas is known as the phase that dominates the central regions, and found to be the most massive outflow component in AGN (see e.g., \citealt{kanekar2008outflowing}; \citealt{cicone2014massive}; \citealt{dasyra2011turbulent}; \citealt{dasyra2012cold}; \citealt{2015A&A...580A...1M}; \citealt{feruglio2010quasar}; \citealt{rupke2013multiphase}; \citealt{garcia2015high}; \citealt{carniani2015ionised}; \citealt{2017A&A...601A.143F}; \citealt{morganti2017many}; \citealt{fluetsch2019cold}), making cold molecular gas the main tracer of AGN outflows.

Several works aim at understanding the origin of non-circular motions, outflow driving mechanisms and the impact of AGN feedback on the interstellar medium (ISM) by imaging and modelling the kinematics of cold molecular gas in the central regions of AGN host galaxies. Both types of non-circular motions (inflows and outflows) are detected in the nuclear and circumnuclear regions of nearby star forming and AGN host galaxies. Using the ALMA observations of dense molecular gas tracers (CO(3-2), CO(6-5), HCN(4-3), HCO+ (4-3), and CS(7–6)) \citet{garcia2014molecular} studied the fueling and the feedback of star formation and nuclear activity in the nearby Seyfert 2 nucleus NGC 1068.
\\
The authors confirm the detection of molecular line and dust continuum emissions from different regions in the source. They also indicate the presence of both an inward radial flow in the starburst ring and the bar region, and a massive outflow, an order of magnitude higher than the star formation rate, in the inner region (r $\sim$ 50 pc out to r $\sim$ 400 pc), interpreting the inward flow as the combined action of the bar and the spiral arms, and the outflow is AGN driven.
Based on the tight correlation between the ionized gas outflow, the radio jet, and the occurrence of outward motions in the disc the autors suggest that the outflow is likely AGN driven.\\
With a better spatial resolution of $\sim$4 pc towards the same galaxy, \citet{2016ApJ...823L..12G} studied the dust emission and the distribution and kinematics of molecular gas by using the ALMA observations of the dust continuum at 432 $\mu$m and the CO(6–5) molecular line emission in its circumnuclear disc. 
They conclude that the overall slow rotation pattern of the disc is perturbed by strong non-circular motions and enhanced turbulence \citep[see also a similar work in][]{2019A&A...632A..61G}.

\citet{2014A&A...565A..97C} report the ALMA observations of CO(3-2) emission in the nuclear region of the Seyfert 1 galaxy NGC 1566. They find a conspicuous nuclear trailing spiral, and weak non-circular motions at the periphery of the nuclear spiral arms. Recently, by analysing the nuclear kinematics of the same galaxy NGC 1566 via ALMA observations of the CO(2-1) emission, \citet{slater2019} show the presence of significant non-circular motions in the innermost (200 pc) and along spiral arms in the central kpc. They find a molecular outflow in the disc with velocities of $\sim 180\, \rm{km\,s^{-1}}$ in the nucleus.

Using the ALMA observations of CO(3-2) emission around the nucleus of NGC 1433, \citet{2013A&A...558A.124C} find also an intense high-velocity CO emission feature redshifted to $200\, \rm{km\,s^{-1}}$, with a blue-shifted counterpart, at 2$^{''}$ (100 pc) from the centre, interpreting the wide component as an outflow partly driven by the central star formation, but mainly boosted by the AGN through its radio jets. Similarly, using the ALMA observations of the infrared-luminous merger NGC 3256, \citet{Sakamoto_2014} detect high-velocity molecular outflows from the northern and southern nucleus with two different large velocities $> 750\, \rm{km\,s^{-1}}$ and [1000$\sim$2000] $\, \rm{km\,s^{-1}}$, respectively, interpreting the northern outflow as a starburst-driven superwind and the southern one as an outflow driven by a bipolar radio jet from an AGN.

In their ALMA CO(2–1) line observations with angular resolutions 0.11$^{''}$-0.26$^{''}$ (9-21 pc), \citet{2018ApJ...859..144A} find strong non-circular motions in the central (0.2$^{''}$-0.3$^{''}$) regions of the nearby Seyfert galaxy NGC 5643 with velocities of up to 110 km/s, explaining the motions as radial outflows in the nuclear disc in the absence of a nuclear bar. \citet{2020A&A...643A.127D} present a detailed analysis of the kinematics and morphology of cold molecular gas in the nuclear/circumnuclear regions of five nearby (19 - 58 Mpc) Seyfert galaxies, Mrk 1066, NGC 2273, NGC 4253, NGC 4388, NGC 7465. The authors detect CO(2-1) emission in all galaxies with disky or circumnuclear ring like morphologies. Moreover, they find in all galaxies, though the bulk of the gas is rotating in the plane of the galaxy, non-circular motions in four of the galaxies (Mrk 1066, NGC 4253, NGC 4388 and NGC 7465). They interpret the non-circular motions in NGC 4253, NGC 4388 and NGC 7465 as streaming motions due to the presence of a large-scale bar, whereas the non-circular motions in the nuclear regions of Mrk 1066 and NGC 4388 are outflows due to the interaction of the AGN wind with molecular gas in the galaxy disc.

Fast and massive molecular outflows are capable to impact the nearby ISM, and molecular mass outflow rate is shown to be well correlated with AGN bolometric luminosity \citep{2017A&A...601A.143F}. \citet{2017A&A...601A.143F} also report that in AGNs with bolometric luminosity up to $\sim$ 10$^{46}$ erg s$^{-1}$ the molecular mass outflow rate dominates its ionised counterpart. Moreover, \citet{2017A&A...601A.143F} find that the molecular gas depletion timescale and the molecular gas fraction of galaxies hosting powerful AGN driven winds are 3-10 times shorter and smaller than those of main sequence galaxies with similar star formation rate, stellar mass, and redshift, indicating that,
at high AGN bolometric luminosity, the reduced molecular gas fraction may be due to the destruction of molecules by the wind, leading to a larger fraction of gas in the atomic ionised phase.

Also using the ALMA observations of CO(2-1), \citet{2015A&A...580A...1M} find that the gas kinematics in the Seyfert 2 galaxy IC 5063 are very complex. The authors detect fast cold molecular gas outflow with velocities up to $650\, \rm{km\,s^{-1}}$, indicating that the outflow can be driven by the central AGN and the radio jet. Similarly, using the CO(2-1) and CO(3-2) ALMA observations of the nearby Seyfert 1.5 galaxy NGC 3227, \citet{2019A&A...628A..65A} show the presence of CO clumps with complex kinematics, dominated by non-circular motions in the central region (1$^{\prime\prime}$ $\sim$73pc). \citet{2020A&A...633A.127F} also show the presence of a prominent jet-driven outflow of CO(2-1) molecular gas along the kinematic minor axis of the Seyfert 2 galaxy ESO 420-G13, at a distance of 340-600 pc from the nucleus.

 Several similar works were published on constraining the kinematics/dynamics of molecular gas in the nuclear/circumnuclear regions of star forming and AGN host galaxies (e.g., \citealt{Tadhunter2014}; \citealt{10.1093/mnras/stz1244}; \citealt{Salak_2020}; \citealt{refId0}; \citealt{comb+19}; \citealt{audibert2019}; \citealt{sirressi2019testing}, see also review by \citealt{Veilleux2020}).\\
In this paper, we present the analysis of the CO molecular gas kinematics in the nuclear regions of three different types of Seyfert galaxies, NGC 4968, NGC 4845, and MCG-06-30-15, by using the ALMA observation of the bright CO(2-1) emission line as a tracer. These galaxies are different in AGN type, AGN luminosity, and morphology. The knowledge about the nature of cold molecular gas kinematics and its origin in the nuclear and circumnuclear regions of such different AGN host galaxies is essential to verify whether the driving mechanism(s) is the same, which in turn could be used as an input to construct a universal model that describes such physical mechanism(s).

The three galaxies belong to the TWIST sample (\citet{2020A&A...633A.127F} and Fern{\'a}ndez-Ontiveros et al, 2021 in prep) which aims at studying 41 AGN extracted from the IRAS 12$\mu m$ flux-limited sample \citep{rush+93} located close enough (10$<$D$<$50 Mpc) to ensure a good spatial resolution through ALMA observations and resolve the morphological structures in the central $\sim$3$\times$3 kpc region, and being above the "knee" of the Seyfert galaxy luminosity function (therefore, being statistically representative to determine the impact of the outflow and inflow mechanisms for the bulk of the AGN population). Moreover the sample has been extensively studied in the past 30 years and have a wealth of ancillary data from the X-rays to the radio (Fern{\'a}ndez-Ontiveros et al, 2021 in prep). 
The three galaxies in this work extend the preliminary study of the TWIST sample initiated in Fern\'andez-Ontiveros et al. (2020) for ESO 420-G13. These include one type 1 and two type 2 nuclei located at ~ 20-40 Mpc, that sample three different bins in X-ray luminosity. Table~\ref{t1} list the properties of the three objects under study.

To study the molecular gas kinematics we use two different softwares, namely the 3D-Based Analysis of Rotating Object via Line Observations $^{3D}$BAROLO \footnote{https://bbarolo.readthedocs.io/en/latest/} (\citealt{teodoro20153d}) and DiskFit (\citealt{peters2017}).
This paper is structured as follows. In Section \S~\ref{pppt} we present a brief description of the properties of the sources, the ALMA observations and data reduction. We discuss modeling of the CO molecular gas kinematics in \S~\ref{mod}. The properties and kinematics of the CO molecular gas, and the residuals in each galaxy are presented in \S~\ref{res} and in \S~\ref{MH2}, \S~\ref{dust}. We discuss the results in \S~\ref{dis}, and the main conclusions in \S~\ref{con}.\\
We adopt a flat $\Lambda$CDM cosmology with $\Omega _{\lambda}$ = 0.7, $\Omega _{\lambda}$ = 0.3, and $H_{0}$ = $70\, \rm{km\,s^{-1}}$ Mpc$^{-1}$. 

\begin{table}
\caption{\label{t1} Physical and geometrical parameters of the galaxies. The SFR is computed from the PAH observations \citep{mor21} and 12$\mu m$ luminosity in [${\rm log_{10} L_{\odot}}$] \citep[from][]{rush+93}. X-ray luminosity from \citet{nanda}}.
\centering
\setlength{\tabcolsep}{2pt}
\begin{tabular}{llccr}
\hline\\
&NGC 4968&NGC 4845&MCG-6-30-15\\
\hline\\
AGN type&S2&S2&S1.2&\\
Redshift&0.00986&0.00411&0.008&\\
Classification&SB0&SABab&Sab&\\
${\rm \log L_{2-10keV}} $ &  43.20    & 41.98      &42.74&\\ 
(erg/s) & & & & \\
${\rm \log L_{12\mu m}}  $ &  9.91    & 9.87      &9.52&\\ 
(L$_{\odot}$) & & & & \\
SFR (M$_{\odot}$/yr) &4.29& - &1.82&\\
Distance&42 Mpc&18 Mpc&37 Mpc&\\
RA&13h07m05.935s&12h58m01.187s&13h35m53.777s&\\
Dec&$-23^{o}40^{'}36.23^{''}$ &$1^{o}34^{'}32.526^{''}$&$-34^{o}17^{'}44.242^{''}$ &\\
INC ($^{o}$) &60&73&[65-68]&\\
PA ($^{o}$) &[234-256]&[62-81]&[115-122]&\\
\hline\\
\end{tabular}\\ \vspace{.05cm}
\end{table}

\section{Basic galaxy properties and observations}\label{pppt}

NGC 4968 is a nearby Seyfert 2 spiral galaxy, classified as a barred Spiral (SB0) according to the Hubble and de Vaucouleurs galaxy morphological classification, located at a redshift of $z$ = 0.00986 (\citealt{lamassa2017chandra}; \citealt{malkan1998hubble}), corresponding to a distance of 42 Mpc.
NGC 4968 Narrow Line Region (NLR) was imaged in the [O III] $\lambda$ 5007 and [N II] $\lambda\lambda$ 6548, 6583 + H$\alpha$ emission lines, as well as in its adjacent continua (centered near 5500 and 8000 {\AA}) using HST by \citet{ferr2000}. From the continuum images these authors %\citet{ferr2000}
estimate a photometric major axis P.A. of 45$^{o}$ and an inclination angle of 60$^{o}$ at radii greater than 10$^{\prime\prime}$ (2.1 kpc) \citep[][see also]{2003ApJS..148..327S}.
The NLR extends toward  the south-east side of the nucleus and the [OIII] line emission has a wedge-shaped morphology filling the cavity inside the dusty ring.
The presence of the dust may be responsible of its morphology but the presence of a ionisation cone projecting against the far side of the  galaxy disc cannot be excluded
\citep{ferr2000}\\
\cite{strong2004molecular} studied the properties of the molecular gas in NGC 4968 with single dish (15-m) millimetre  observations at the ESO-SEST telescope. They estimate the line luminosity (${\rm L_{CO}}$) and molecular mass (${\rm M({H_2})}$) through observations from CO(1-0) (6.0$\times$10$^{7}$ K km s$^{-1}$pc$^{2}$ and $21.0\times10^{7}$ M$_{\odot}$, respectively), and CO(2-1) (4$\times$10$^{7}$ K km s$^{-1}$pc$^{2}$ and $15.0\times10^{7}$ M$_{\odot}$, respectively), making use of  a CO-to-H$_{2}$ conversion factor, $\alpha_{\rm CO}$, of 3.47.

%NGC 4968: this Sy 2 galaxy has been imaged in[OIII]and[NII]usingHSTby Ferruit et al.(2000)and in[OIII]bySchmitt et al.(2003). WithHST, there is a fan-shaped ENLRextended toward the SE by∼1.5 arcsec. This can be traced outto 7 arcsec in our data, and the counter ionization cone is alsofaintly visible. Previous ground-based data by Pogge(1989)did not detect any extended[OIII]emission.

The galaxy NGC 4845 is a nearby Seyfert 2 with a clear starburst component \citep{Thomas}, classified as an intermediate barred spiral galaxy (SABab), at $z$ = 0.00411 located in the Virgo Southern Extension (\citet{Irwin_2015} and references therein). We adopt the Tully-Fisher distance of 18 Mpc. The inclination on the sky is almost edge-on, revealing contrasted dust lanes on the near side, and a peanut-shape for the bulge. The galaxy contains a bright unresolved core with a surrounding weak central disc (1.8 kpc diameter). The radio spectrum of the core has been known to evolve with time, which could be due to an adiabatic expansion (outflow), likely in the form of a jet or cone (\citealt{Irwin_2015}). Using spectral observations of NGC 4845 in different position angles (PA = 44$^{o}$, 78$^{o}$, 98$^{o}$, 118$^{o}$, 178$^{o}$), \cite{bertola1989evidence} studied the kinematics of the ionised gas in the central region (r $\leq$ 1.5 kpc) and revealed a regular but non-axisymmetric velocity field. Based on photometry, \cite{bertola1989evidence} also point out the presence of a possible slight twisting between the disc and bulge isophotes, interpreting this as an indication of triaxial bulge in NGC 4845.
This galaxy shows a ionisation cone with an opening angle of 120$^o$ perpendicular to the circumstellar disc of HII regions, the counter inionisation cone is faint, likely because of exinction \citep{Thomas}.

The early-type Sab galaxy MCG-06-30-15 is a Seyfert 1.2 galaxy%\footnote{http://leda.univ-lyon1.fr/} 
(\citealt{2014A&A...570A..13M}) located at a distance of 37 Mpc, $z$ = 0.008 (\citealt{2009ApJ...701..658W}).
This active galaxy has a 400 pc diameter stellar kinematically distinct core (KDC) counter-rotating with respect to the main body of the galaxy (\citealt{raimundo2016tracing}). The molecular gas, traced by the H$_2$ 2.12 $\mu$m emission is also counter rotating with respect to the main stellar body of the galaxy, implying that the formation of the distinct core is associated with inflow of external gas into the centre of MCG-6-30-15 and the event that formed the counter-rotating core is also the main mechanism providing gas for the AGN fuelling. Moreover, this shows that external gas accretion is able to significantly replenish the fuelling reservoir in such active galaxies. MCG-06-30-15  NLR was imaged using HST by \citet{ferr2000} in the [O III] $\lambda$ 5007 and [N II] $\lambda\lambda$ 6548, 6583 + H$\alpha$ emission lines, as well as in their adjacent continua (centered near 5500 and 8000 {\AA}). \citet{ferr2000} estimate the photometric major axis P.A. and an inclination angle of 115$^{o}$ and 60$^{o}$, respectively. The [O III] image reveals a nuclear extension aligned parallel to the photometric major axis of the galaxy, which presumably represents gas coplanar with the stellar disc \citep[see also][]{2003ApJS..148..327S}.\\
\citet{Rosario} measure the CO(2-1) single dish flux and discuss the properties of its molecular gas in relation to its AGN properties. These authors estimate the line luminosity (${\rm L_{CO}}$) and molecular mass (${\rm M({H_2})}$) through observations from CO(2-1) (1.3$\times$10$^{7}$ K km s$^{-1}$pc$^{2}$ and $1.4\times10^{7}$ M$_{\odot}$, respectively) using a CO-to-H$_{2}$ conversion factor, $\alpha_{\rm CO}$, of 1.1.

\subsection{ALMA observations and data reduction}\label{obs}
\subsubsection{ALMA observations}
The observations of the bright CO(2-1) line at 230.5 GHz rest frequency (in Band 6) were carried out as part of ALMA cycle 5 under project IDs 2017.1.00236.S in December 2017 and January 2018. With a spatial resolution between 25pc and 48pc, the CO(2-1) maps are sensitive enough to detect molecular masses as low as $\sim$ 10$^{5}$ M$_{\odot}$ (5$\sigma$) per beam. See the summary of ALMA observations in Table \ref{t1}.

\subsubsection{Data reduction and imaging}
Data were calibrated and post-processed using the Common Astronomy Software Applications (CASA) package \citep{2007ASPC..376..127M}, applying the standard calibration recipes provided by the ALMA Observatory. The data were calibrated using the \textsc{casa} version 4.7.0 and the calibration script provided by the observatory, while further post-processing was done using the CASA version 5.4.0. In all cases, the \textit{hogbom} deconvolution algorithm was applied using \textit{briggs} weighting and a robustness value 2.0 to reconstruct the final datacubes, optimising the value of the limiting flux threshold in each case. The spectral datacubes for the emission lines were produced with a channel width of $\sim 10$ and
$\sim 30\, \rm{km\,s^{-1}}$ and a pixel size of about 1/5$^{th}$ to 1/4$^{th}$ of the synthesised beam size (see Table \ref{t1} for the synthesised beam size). The emission line regions were automatically masked during the cleaning process in the spectral cubes using the "auto-multithresh" algorithm in \textit{tclean}.\\
The continuum emission was then subtracted in the spatial frequency domain - i.e. prior to the image reconstruction - using a 1D polynomial interpolation between the adjacent continuum channels at both sides of the respective emission lines
Continuum maps were constructed using all the continuum channels in the four spectral windows, that is discarding the channels with CO(2-1) emission. The masking procedure for the continuum data was run interactively during the cleaning process.\\
 Finally, all datacubes were corrected for the primary beam attenuation pattern. Using the final CO(2-1) datacubes, the first three moments corresponding to the integrated intensity map of the line, the average velocity field and the average velocity dispersion map were computed numerically. In order to reduce the noise in the moment maps, for each spaxel only those channels with a signal-to-noise (S/N) above $\sim$5 the median absolute deviation were considered, following the approach in \citet{2020A&A...633A.127F}.
\begin{table}
\caption{\label{t2} ALMA CO(2-1) observation log}
\centering
\setlength{\tabcolsep}{1.8pt}
\begin{tabular}{lllll}
\hline\\
&Target line CO(2-1)\\
\hline\\
&NGC 4968&NGC 4845&MCG-06-30-15\\
\hline\\
%\hline\\
$\nu_{rest}$ & 230.5 GHz&230.5 GHz&230.5 GHz&\\
Date & Dec 2017&Dec 2017&Dec 2017&\\
&Jan 2018&Jan 2018&Jan 2018&\\
Array configuration & C43-5 &C43-5&C43-5&\\
Spatial resolution & 48pc&45pc&25pc&\\
Channel width &2.5 km s$^{-1}$ &2.5 km s$^{-1}$&2.5 km s$^{-1}$&\\
Rms sensitivity &1.4 mJy/beam& 7 mJy/beam&0.7mJy/beam&\\
Synthesized beam &0.27$^{''}$x0.22$^{''}$ &0.56$^{''}$x0.48$^{''}$&0.16$^{''}$x0.13$^{''}$&\\
\hline\\
\end{tabular}\\ %\vspace{.05cm}
\end{table}

\section{Modeling the kinematics of CO(2-1) emission line}\label{mod}
\subsection{3D modeling of the main rotating disc}
We use the CO(2-1) emission line as a tracer of the main kinematic features of the CO gas in the nuclear region of the galaxies. To constrain the gas kinematical perturbations we construct a 3D disc model using the 3D-Based Analysis of Rotating Object via Line Observations ($^{3D}$BAROLO) software and fit the model to the CO (2-1) emission line datacube. This software  automatically fits 3D tilted-ring models to emission-line data cubes. The model assumes that  the emitting material at each radius is confined within a geometrically thin disc and its kinematics are dominated by pure rotational motion (\citealt{teodoro20153d}).
The model requires the geometrical parameters of the disc, namely the centre position coordinates, the inclination angle (INC) and the kinematic position angle (PA). When the emission peaks at the centre, like in the case of NGC4845, the coordinates can be determined by fitting a 2D Gaussian to the central part (where the peak emission is) of the intensity map (see \citealt{sirressi2019testing}), whereas for maps without a central emission peak (the case of NGC 4968 and MCG-06-30-15) the code uses a source-finder algorithm to identify the emission region, and then calculates the geometrical centroid, weighted by the flux intensity.
The centre positions are determined and reported in Table \ref{t2}. The INC, the angle between the normal to the disc axis and the line of sight, can be inferred from the 0$^{th}$ moment map or from the 1$^{st}$ moment map depending on how well these moments can be estimated. For NGC 4845 we determine the INC parameter from the ratio of the two axes, major and minor axes, in the 1$^{st}$ moment (velocity) map (see e.g. \citealt{sirressi2019testing}). But for well resolved data (the case of NGC 4968) it is better to use the 0$^{th}$ moment (intensity) map rather than the velocity field. The inclination angle INC turns out to be 60$^{o}$ (NGC 4968) by fitting an ellipse to the outer ring in the intensity map. For NGC 4845, the INC = 73$^{o}$ is determined from the velocity field. For MCG-06-30-15, 3D-Barolo estimates that the INC angle varies from 65$^{o}$ to 68$^{o}$. \\
3D-Barolo measures the PA as the angle between the North and the receding half of the major axis (= positive velocities in the velocity field), measured counterclockwise (see \citealt{teodoro20153d}). The PA varies from 234$^{o}$ to 256$^{o}$ (NGC 4968) and from 62$^{o}$ to 81 $^{o}$ (NGC 4845). For MCG-06-30-15, 3D-Barolo estimates that the PA varies from 115$^{o}$ to 122$^{o}$. 

Using the estimated geometrical parameters and assuming a number of rings at different radial separation we construct a disc model by using the 3D-Barolo kinematic model and fit its emission to the continuum-subtracted ALMA CO(2-1) line datacube.
We use the ALMA CO(2-1) line datacube with a channel widths of $10\, \rm{km\,s^{-1}}$. \\
The number of rings can be determined from the spatial extension of the object. The model uses 10 rings with a radial separation of 0.217 arcsec.
All rings are placed at $N*RADSEP + RADSEP/2$, where $N$ is the number of rings and $RADSEP$ is the radial separation in arcsec. 

Since 3D-Barolo measures the noise at the spectral channel edges when it builds the mask, to avoid the noise effect at the edges, instead of using the continuum-subtracted ALMA CO(2-1) line datacube with velocity range = $(-1000, 990)\, \rm{km\,s^{-1}}$, we consider the central regions containing the emission of the galaxies, reducing the velocity range of the input NGC 4968 datacube to $(-340, 340)\, \rm{km\,s^{-1}}$. To construct the disc model for NGC 4968, we use 27 rings with a radial separation of 0.12 arcsec. \\
The velocity range of the input NGC 4845 datacube is reduced to $(-510, 240)\, \rm{km\,s^{-1}}$ and we use 22 rings with a radial separation of 0.55 arcsec. 
The Barolo fits for N4845 for the velocity and the velocity dispersion are not optimal, however the results do not improve letting the inclination angle to vary between 73 and 80 degrees.

For MCG-06-30-15, we feed the continuum subtracted line datacube to the model without reducing its velocity range. The model is able to find the best model reducing the residuals at minimum.
This could be due to the fact that the kinematics is purely dominated by regular rotation pattern (i.e. a simple kinematics). A slight deviation appears between the blue and red contours in p-v diagramme.

% did not appear for the 30 km/s datacube (you may look at the old draft versions), and this small deviation is indeed visible in the residual profile line 
The Barolo fits for N4845 for the velocity and the velocity dispersion are not optimal, however the results do not improve letting the inclination angle to vary between 73 and 80 degrees.

\begin{figure}
   \centering
   \includegraphics[width=0.4\textwidth,angle=0]{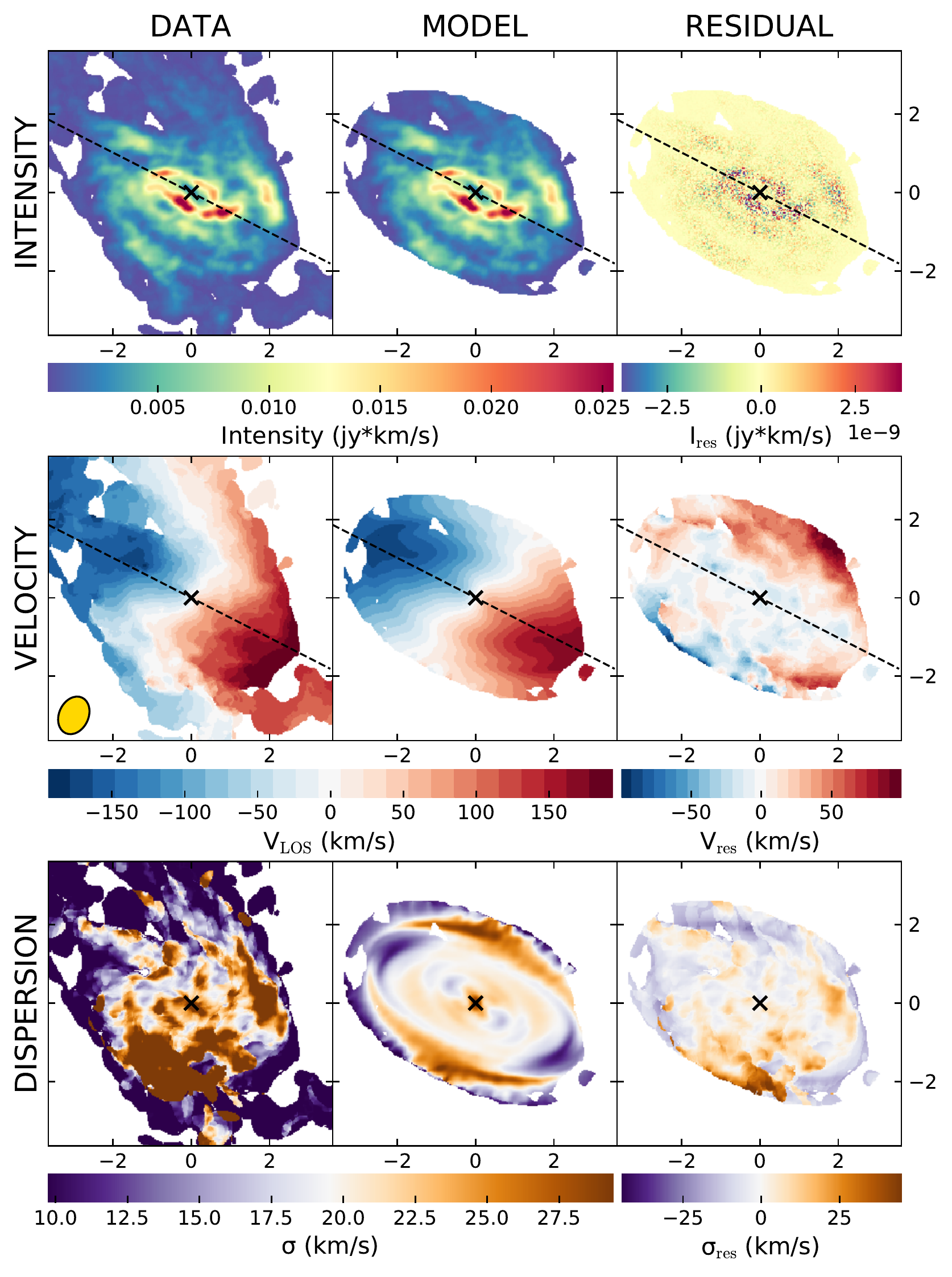}
   \includegraphics[width=0.4\textwidth,angle=0]{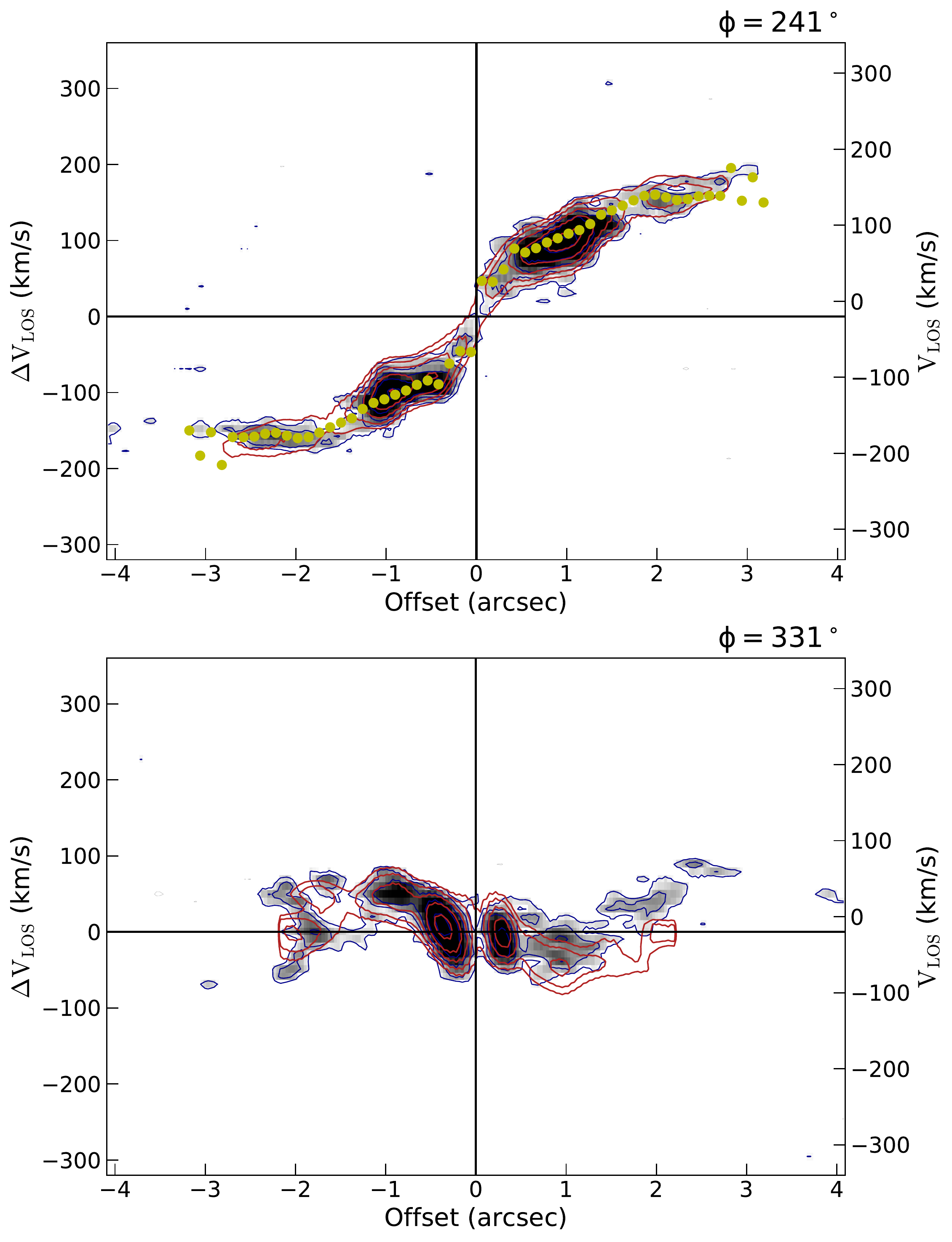}
   \caption{{\bf Upper panels: }\textit{left panels (top to bottom)}: the $0^{th}$ (intensity), $1^{st}$ (velocity field) and $2^{nd}$ (velocity dispersion) moment maps of the ALMA CO(2-1) data of NGC 4968. \textit{central panels}: the same as the left panels but for the best-fit model constructed with $^{3D}$BAROLO. \textit{right panels}: same as the left panels but for the residual (data-model). A black cross marks the centre of the galaxy. The major axis of the molecular disc is shown by a black-dashed line in the velocity map. The yellow ellipse in the bottom-left corner of the data velocity map shows the synthesized beam size (0.27$^{''}$x0.22$^{''}$) with P.A = 245$^{o}$. North and East directions correspond to top and left, respectively. {\bf Lower panels: }: The $p-v$ diagrams extracted from the data-cube (blue solid contours) and model-cube (red solid contours) along the major axis (top panel) and along the minor axis (bottom panel). The contours level of both the data and the model are at [1,2,4,8,16,32,64]*$l$, where $l$ = 0.0012. The rotation velocity of each ring of the best-fit disc model is represented by the yellow solid dots in the top panel.}
      \label{Barolo-4968}        
    \end{figure}

\begin{figure}
   \centering
   \includegraphics[width=0.4\textwidth,angle=0]{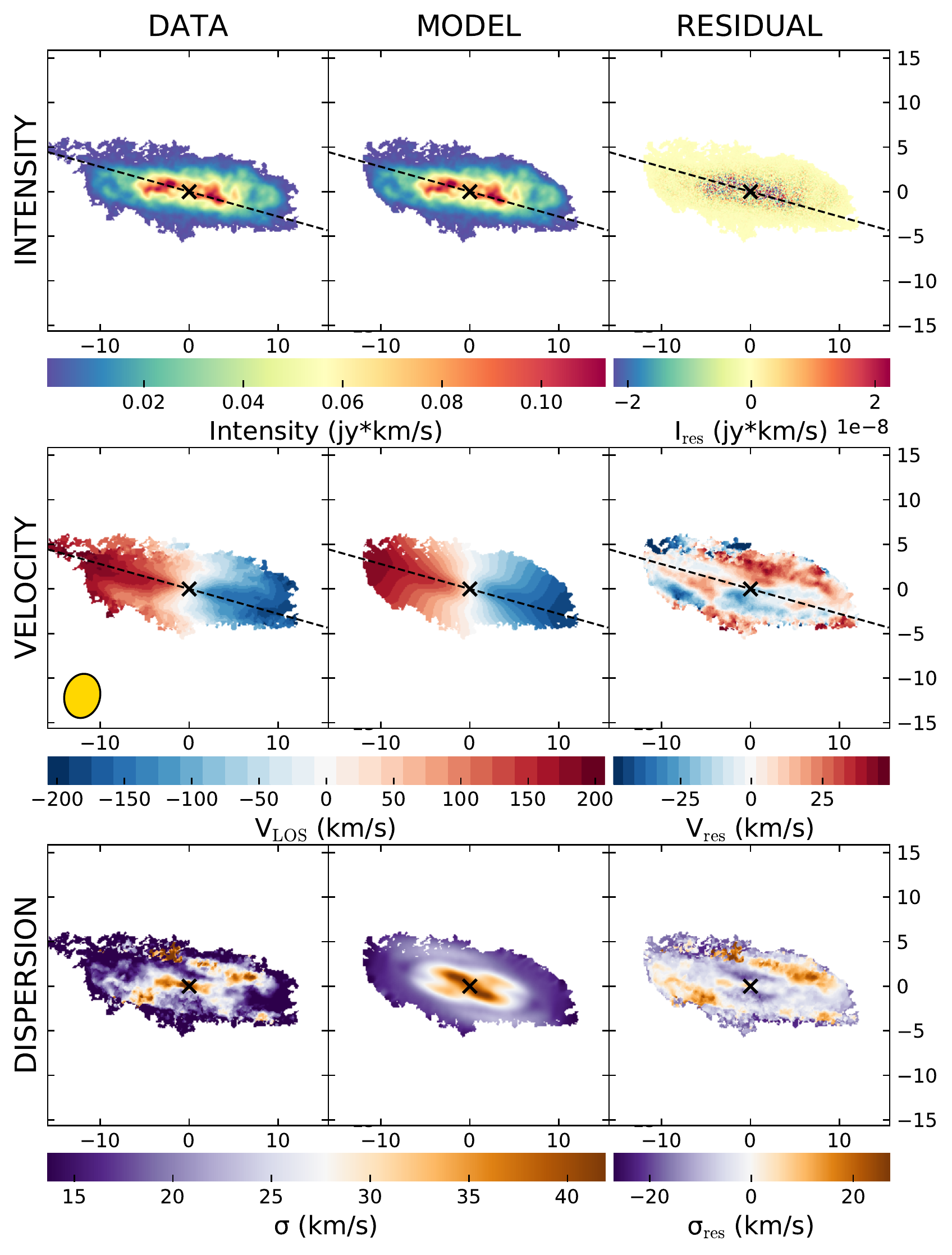}
   \includegraphics[width=0.4\textwidth,angle=0]{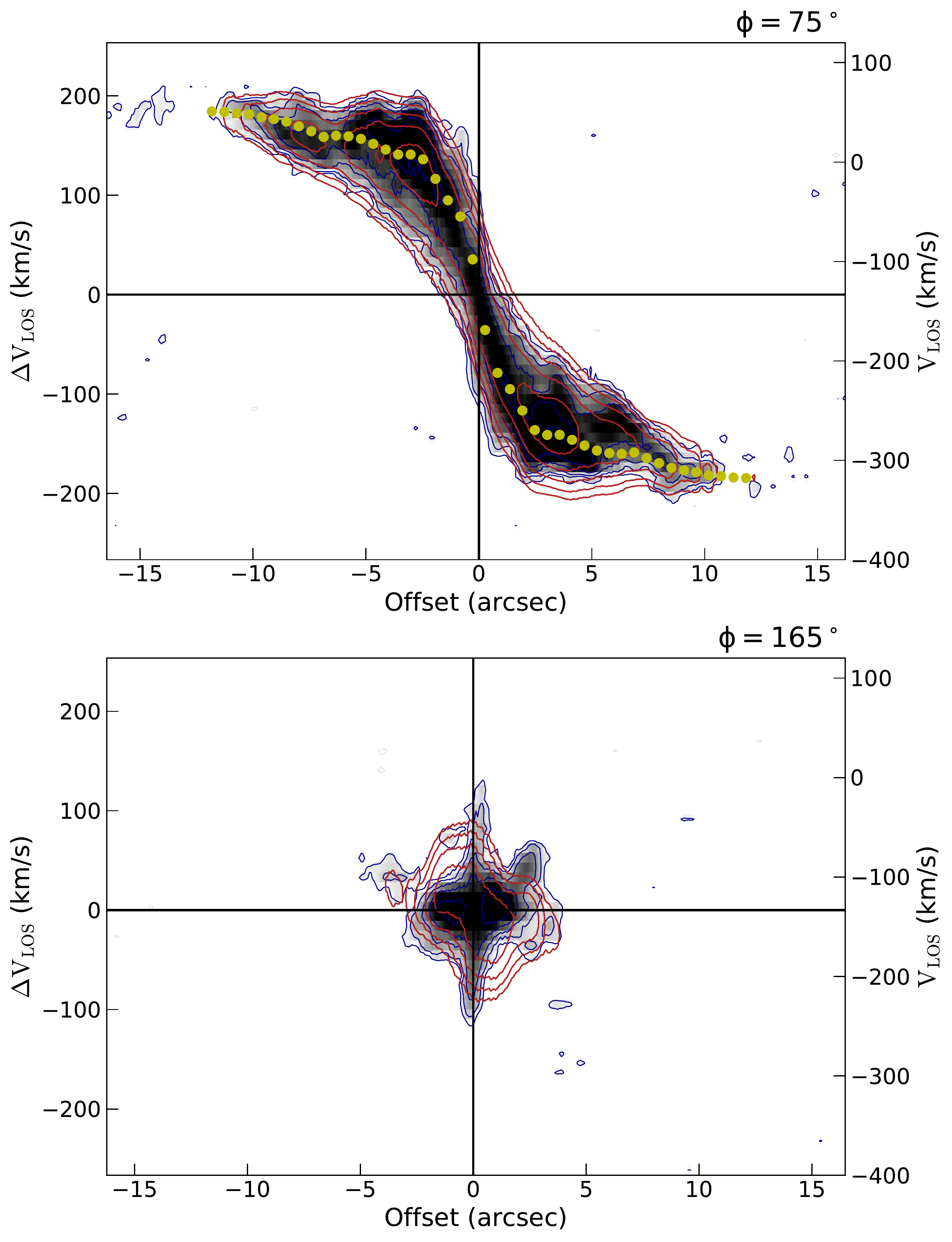}
   \caption{The same as in Fig. \ref{Barolo-4968} but for NGC 4845. The synthesized beam size (0.56$^{''}$x0.48$^{''}$) is plotted in yellow in the bottom-left corner of the data velocity map with P.A = 77$^{o}$. The contours level of both the data and the model are at [1,2,4,8,16,32,64]*$l$, where $l$ = 0.00393635.}
   \label{Barolo-4845}
\end{figure}

\begin{figure}
   \centering
   \includegraphics[width=0.4\textwidth,angle=0]{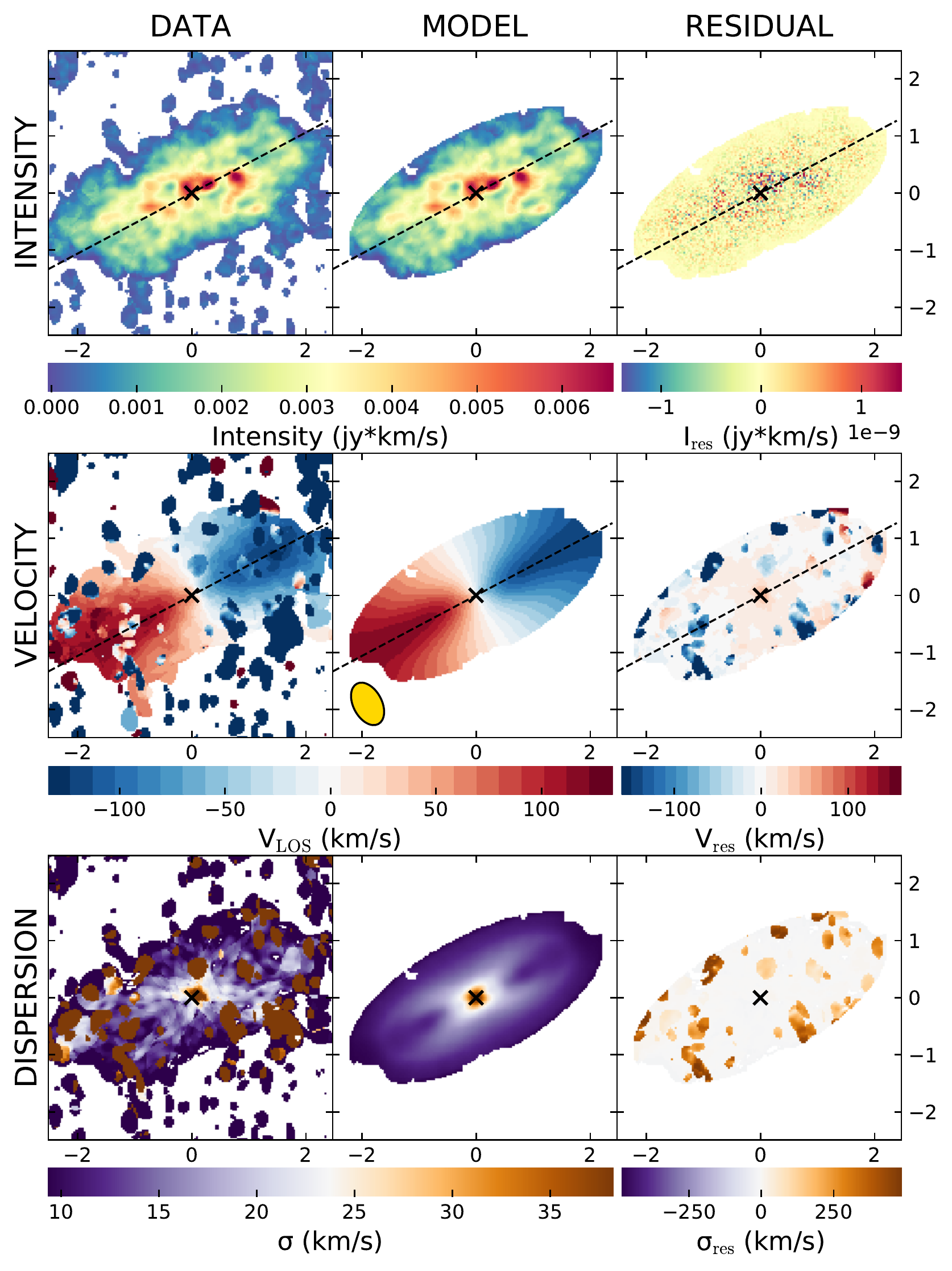}
   \includegraphics[width=0.4\textwidth,angle=0]{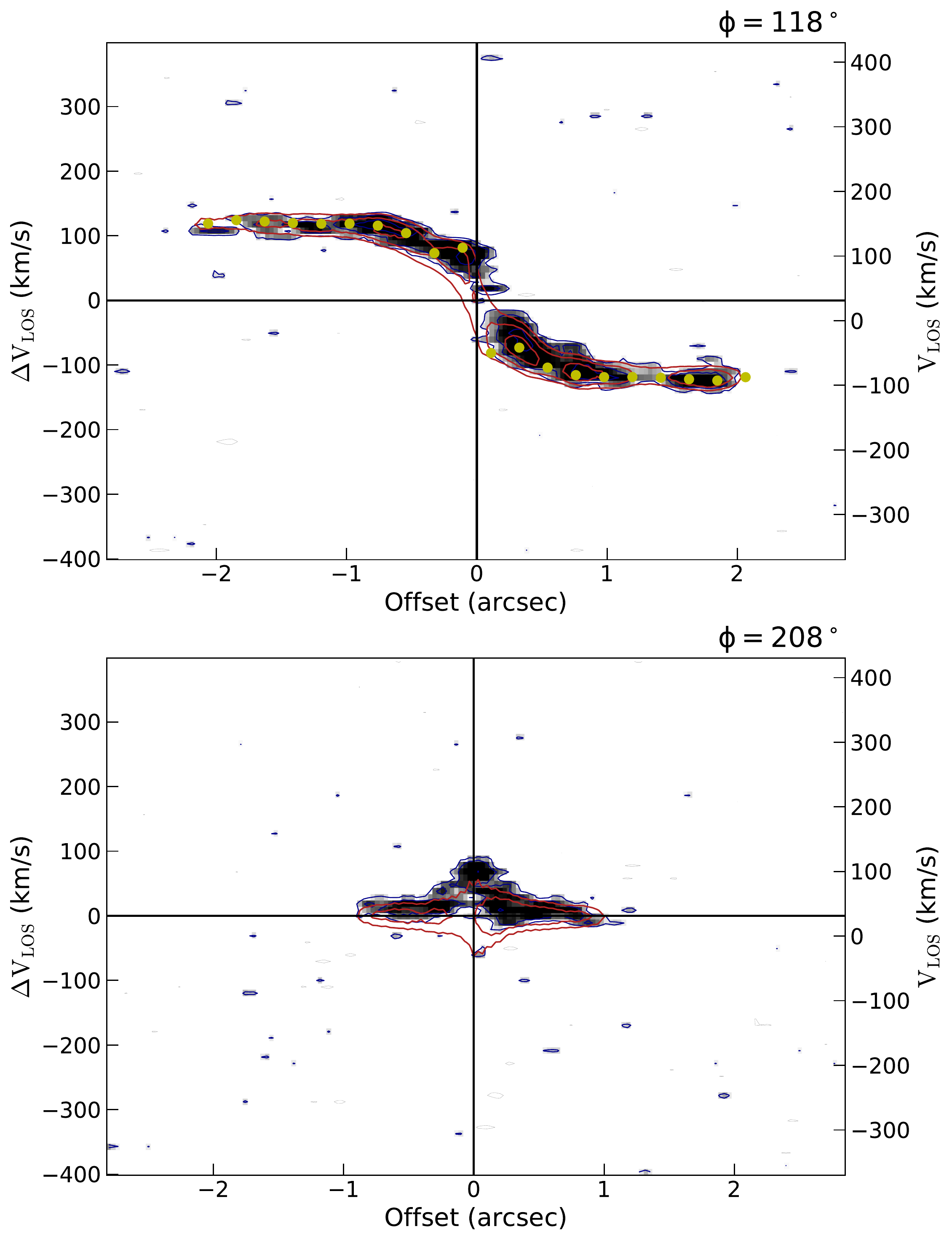}
   \caption{The same as in Fig. \ref{Barolo-4968} but for MCG-06-30-15. The synthesized beam size (0.16$^{''}$x0.13$^{''}$) is plotted in yellow in the bottom-left corner of the model velocity map with P.A = 118$^{o}$. The contours level of both the data and the model are at [1,2,4,8,16,32,64]*$l$, where $l$ = 0.00120941.}
   \label{Barolo-mcg}
\end{figure}

The results from $^{3D}$BAROLO, the kinematic maps and $p-v$ diagrams, are given in Figs. \ref{Barolo-4968}, \ref{Barolo-4845} and \ref{Barolo-mcg} for NGC 4968, NGC 4845, and MCG-06-30-15, respectively. The intensity maps of NGC 4968 and MCG-06-30-15 reveal the molecular gas distribution in the nuclear regions as a ring-like morphology. The gas distribution in the NGC 4845 appeared to have an edge-on inclination.

\subsection{DiskFit modeling}
The DiskFit software fits simple axisymmetric and non-axisymmetric non-parametric models either to photometric images or to kinematic maps  (velocity fields) of disc galaxies. DiskFit fits an entire velocity field with a physically motivated model. The underlying assumption is that the circular orbit of a region of gas is affected by higher order perturbations (i.e. "harmonics"). Physically, $m$ = 1 harmonics correspond to "lopsided" perturbations and $m$ = 2 harmonics correspond to bisymmetric (i.e. bar) perturbations. The model is given by \citet{peters2017} :
\begin{equation}
V_{m}  = V_{s} + \sin i [V_{t}\cos (\theta) - V_{m, t}\cos (\theta_{b})\cos (\theta) - V_{m, r}  (\sin 2\theta_{b})\sin (\theta)]
\end{equation}
where $V_{m}$ is the model (with $m$ specifying the harmonic order with $m$ = 1 or $m$ = 2 in the disc plane), $V_{s}$ is the systemic velocity, $V_{t}$ is the mean orbital speed, $\theta$ is the angle between a point in the disc relative to the major axis, $\theta_{b}$ is the angle between a point in the disc relative to the bar, $V_{m, t}$ and $V_{m, r}$ are the tangential and radial components of the non-circular motions, $\theta$ and $\theta_{b}$ are the azimuthal angles relative to the major axis and the non-circular flow axis, respectively, and $i$ is the disc inclination.
%If $m$ = 1, the model describes a lopsided flow; if $m$ = 2 the model is bisymmetric and describes a barred or elliptical flow.
DiskFit can also fit radial flows (with $m$ = 0 distortions to the flow in the disc plane) and symmetric warps in the outer disc.

DiskFit requires a flat inner disc, but it allows for a symmetric warp in the outer disc. The disc is assumed to be flat out to a warp radius $r_{w}$, beyond which both the ellipticity and the position angle of the line of nodes vary in proportion to $(r - r_{w})^{2}$. In addition, DiskFit can find the best-fitting inner warp radius $r_{w}$ and peak change in ellipticity and position angle of the line of nodes, or it can hold any combination of these parameters fixed. For an axisymmetric model, DiskFit estimates the circular speed for kinematic data. For non-axisymmetric model, DiskFit provides quantitative estimates of the non-circular flow speeds and an estimate of the mean circular speed when run on velocity fields.

% 3D-BAROLO RESULTS FOR NGC 4968    

% 3D-BAROLO RESULTS FOR NGC 4845 [10km/s]

% 3D-BAROLO RESULTS FOR MCG-06-30-15 [10 km/s]

\begin{figure}
  \centering
 \includegraphics[width=0.55\textwidth,angle=0]{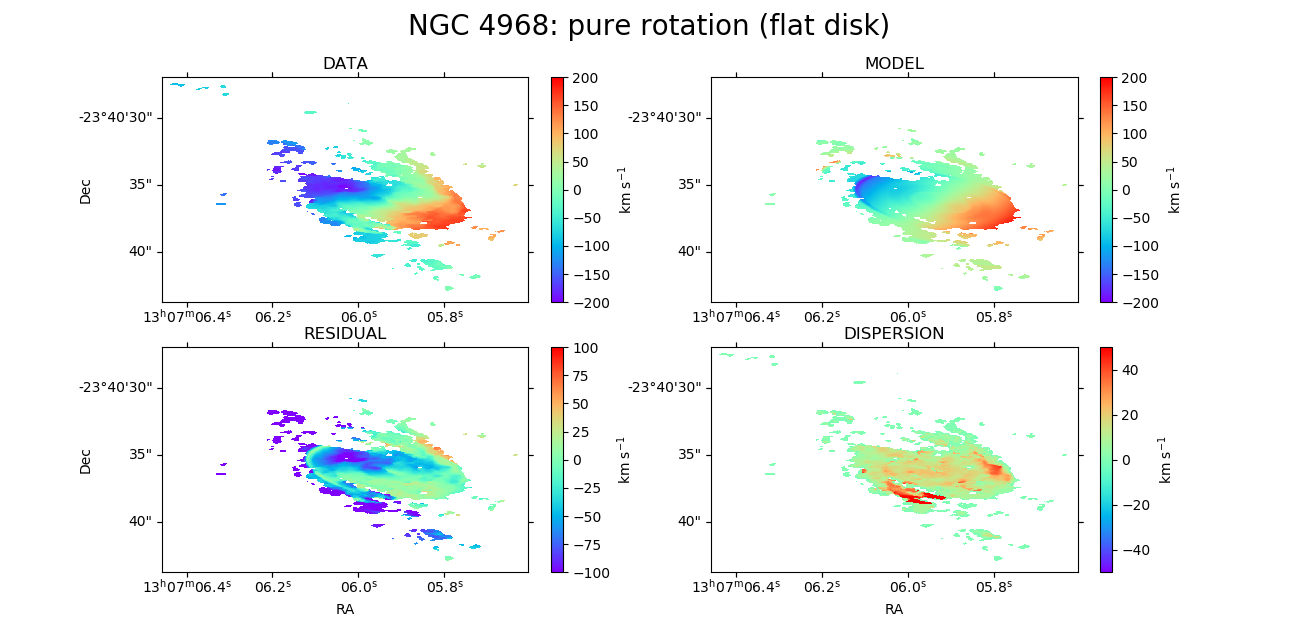}
 \includegraphics[width=0.55\textwidth,angle=0]{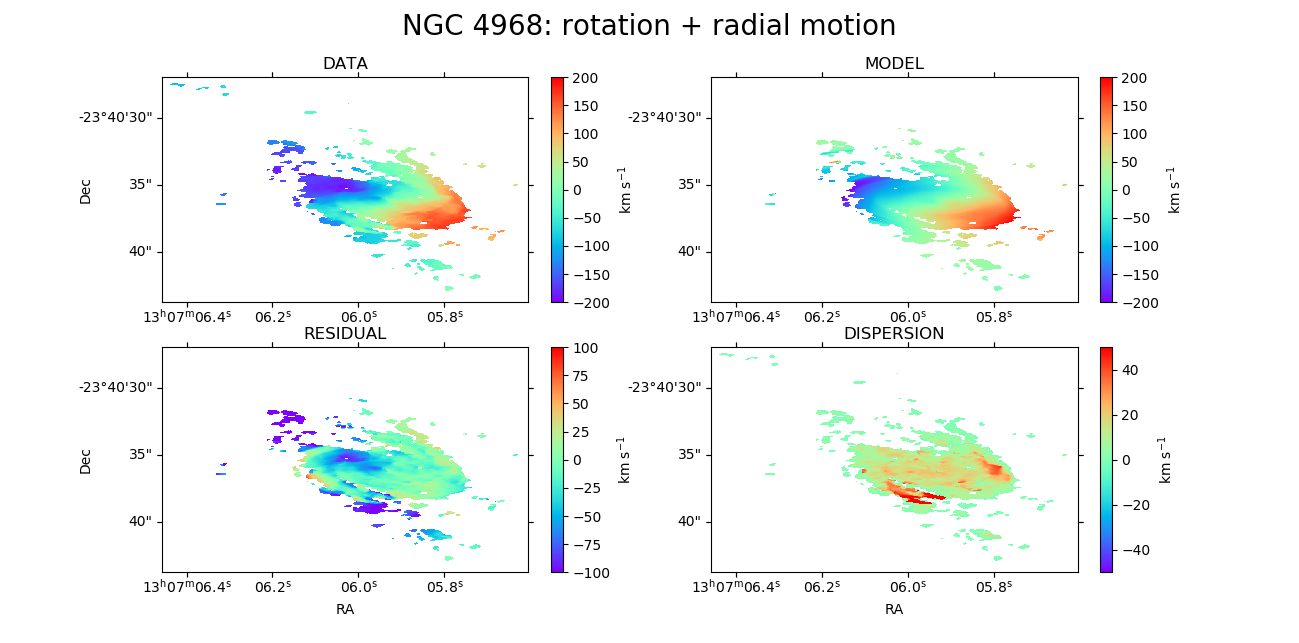}
 \includegraphics[width=0.55\textwidth,angle=0]{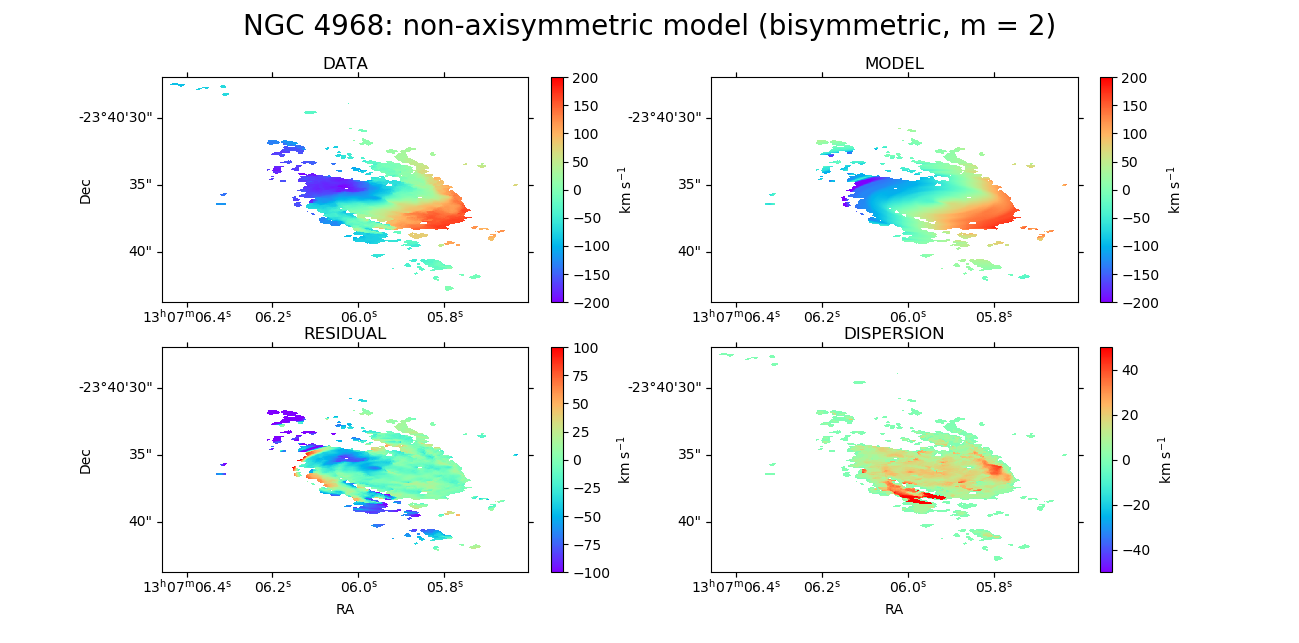}
 \caption{Diskfit results for NGC 4968. \textit{Top panels}: Pure rotation DiskFit model(flat disc). \textit{Middle panels}: Rotation + radial motion DiskFit model. \textit{Bottom panels}: Bisymmetric DiskFit model. For every panel the top left figure reports the data, the top right the model, the low left the residual obtained subtracting the model from the data and the low right figure the dispersion.}
 \label{Diskfit-4968}
\end{figure}

\begin{figure}
  \centering
  \includegraphics[width=0.55\textwidth,angle=0]{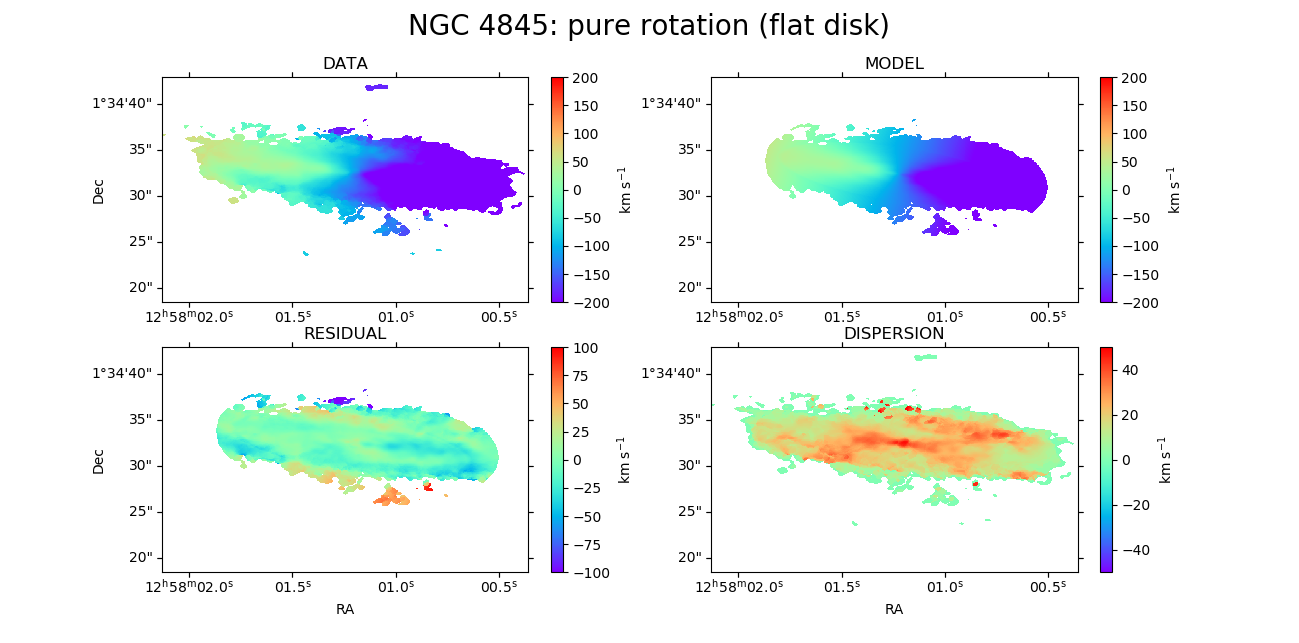}
  \includegraphics[width=0.55\textwidth,angle=0]{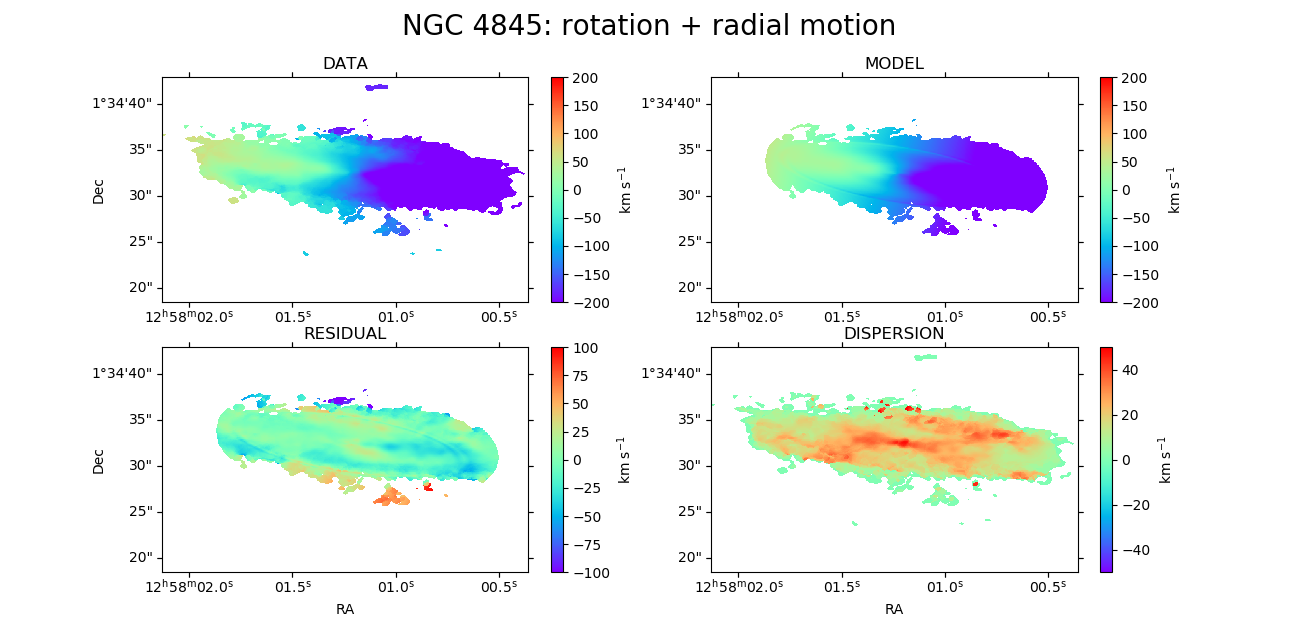}
  \includegraphics[width=0.55\textwidth,angle=0]{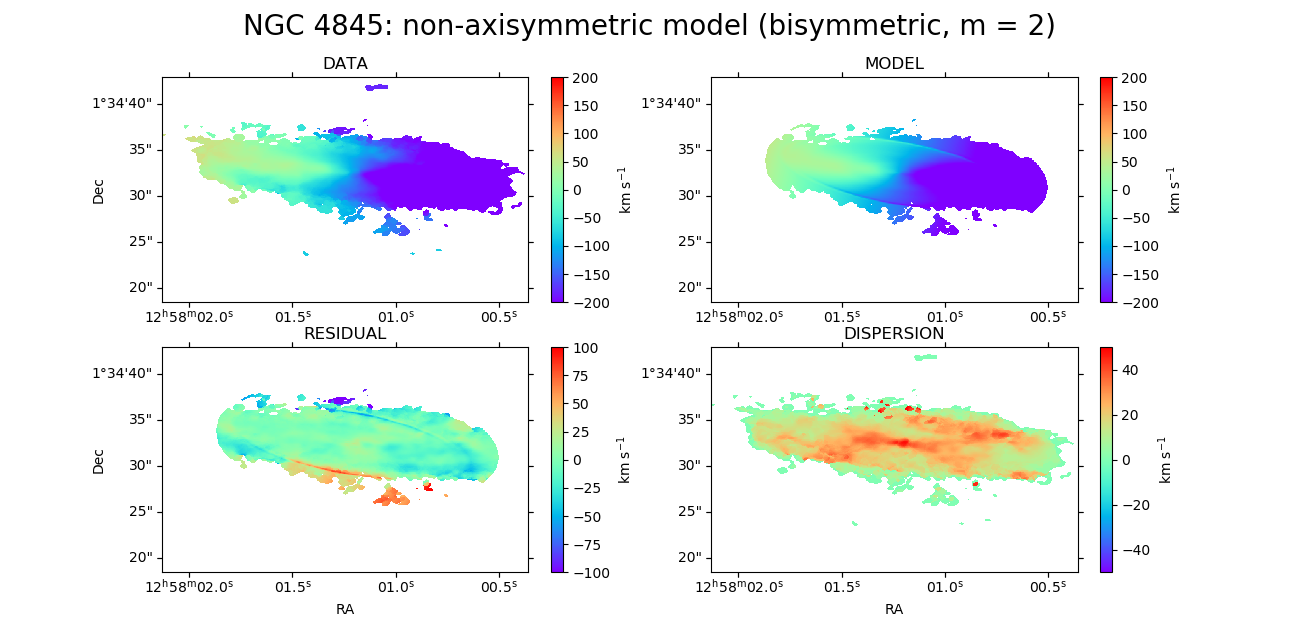}
  \caption{DiskFit results for NGC 4845. \textit{Top panels}: Pure rotation DiskFit model (flat disk). \textit{Middle panels}: Rotation + radial motion DiskFit model. \textit{Bottom panels}: Bisymmetric DiskFit model. For every panel the top left figure reports the data, the top right the model, the low left the residual obtained subtracting the model from the data and the low right figure the dispersion.}
 \label{Diskfit-4845}
\end{figure}

\begin{figure}
 \centering
 \includegraphics[width=0.55\textwidth,angle=0]{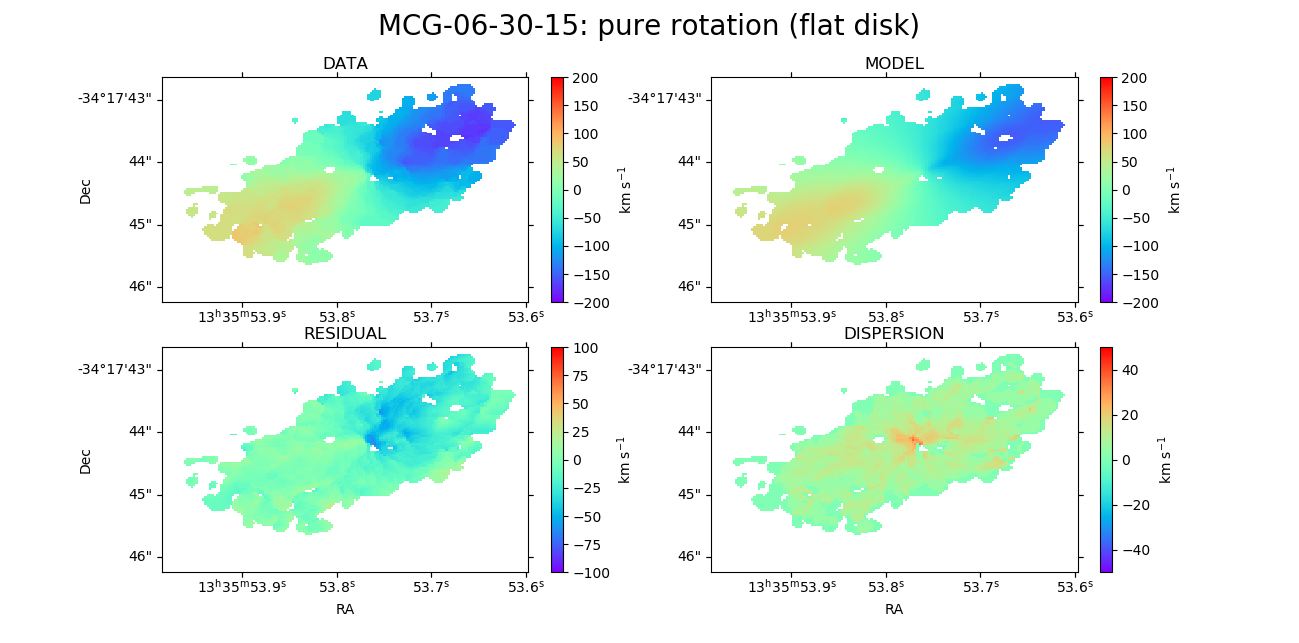}
 \includegraphics[width=0.55\textwidth,angle=0]{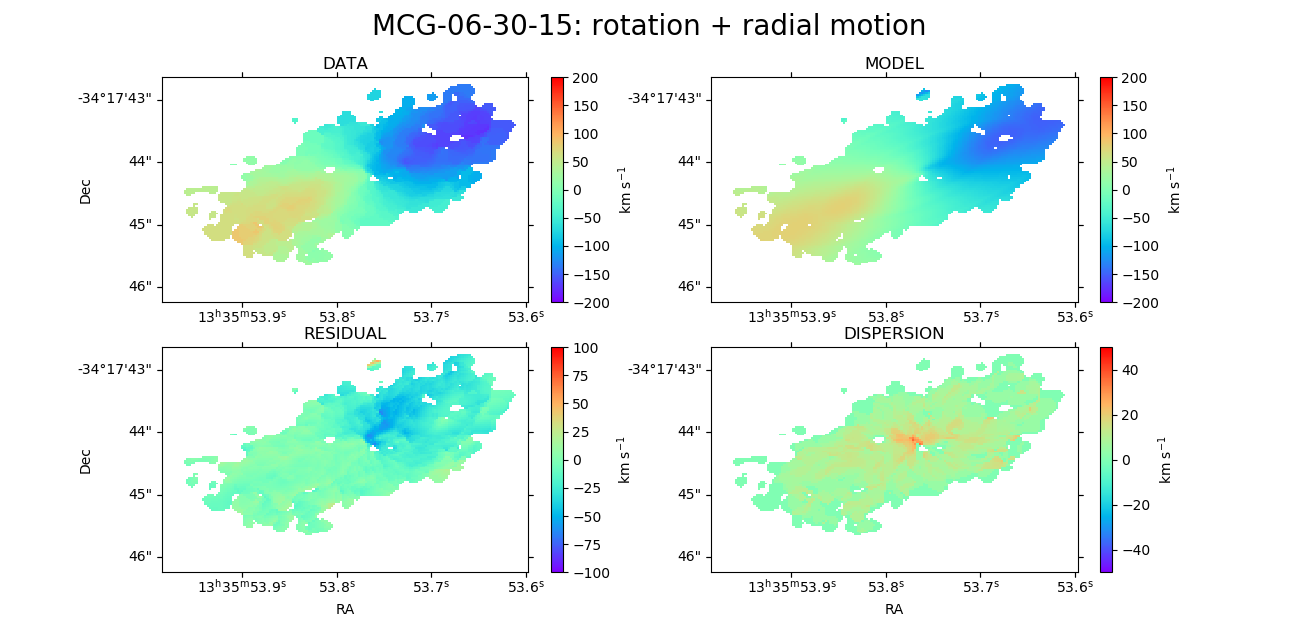}
   \caption{DiskFit results for MCG-06-30-15. \textit{Top panels}: Pure rotation DiskFit model (flat disc). \textit{Bottom panels}: Rotation + radial motion DiskFit model. For every panel the top left figure reports the data, the top right the model, the low left the residual obtained subtracting the model from the data and the low right figure the dispersion.}
 \label{Diskfit-mcg}
\end{figure}

Different axisymmetric and non-axisymmetric DiskFit kinematic models, such as pure rotation (flat disc), pure rotation (flat disc with warp), rotation plus radial motion, lopsided, and bisymmetric models are shown in Figs. \ref{Diskfit-4968}, \ref{Diskfit-4845}, and \ref{Diskfit-mcg} for NGC 4968, NGC 4845, and MCG-06-30-15, respectively.

\section{The CO(2-1) residual emission}\label{res}
%\subsection{The kinematics of the CO molecular gas in the disc}\label{ppt}

% DISKFIT RESULTS FOR NGC 4968 [10 km/s] 

% DISKFIT RESULTS FOR 4845 [10 km/s]

% DISKFIT RESULTS FOR MCG-06-30-15 [10 km/s]

In the following sections, we present the properties and kinematics of the residual (both from $^{3D}$BAROLO and DiskFit) CO molecular gas emission.

\subsection{Residuals from $^{3D}$BAROLO}\label{resB}
As explained in Sec. \ref{mod}, 3D-Barolo makes a tilted-ring rotating disc model of the input data, hence if there are non-circular motions, such as outflows, the model will not be able to fully reproduce the observations. The non-circular motions could be misinterpreted as tilted rings of the plane. Some non-circular motions can be isolated from the rotating disc simply by subtracting the regular rotation pattern from the observed kinematics (observation-rotating disc model). \\

To see whether there are significant residuals in the CO(2-1) motions, we calculate the residuals in the kinematic maps (see Fig. \ref{Barolo-4968} for NGC 4968, Fig. \ref{Barolo-4845} for NGC 4845, and Fig. \ref{Barolo-mcg} for MCG-06-30-15) by subtracting the 3D datacube produced by the model from the observations (the input continuum-subtracted CO(2-1) emission line datacube).

The $^{3D}$BAROLO model reproduces the observations of NGC 4968 relatively well along the kinematic major axis, leaving important residuals mainly in the north-east direction of the kinematic minor axis, as revealed by the velocity map in Fig. \ref{Barolo-4968}.
The residuals reach a maximum velocity %($v-v_{sys}$)
of $\sim 90\, \rm{km\,s^{-1}}$, that is blueshifted to the south and south-east of the minor axis %with velocity $\sim 70\, \rm{km\,s^{-1}}$
 and redshifted to the north and north-west of the minor axis with velocity $\sim 160\, \rm{km\,s^{-1}}$.
%The residual datacube shows $\sim -90\rm {km\,s^{-1}}$ (blue shifted) and  $\sim 160\rm{km\,s^{-1}}$ (redshifted). % The systematic velocity of this galaxy is -20 km/s (as best estimated by BAROLO).  What I did next is that I subtracted this systematic velocity from the blue shifted component as [90 -20] km/s = 70 km/s, which is the real blue shifted comp. of the gas. Why real?  I considered that 20 km/s of the blue shifted component was due to the motion of the galaxy itself rather than the gas within it (systematic velocity).
Therefore, the maximum residual velocity is the mean of the two components as we read them from the residual datacube $\sim$ 125 $\rm{km\,s^{-1}}$.

% Note that: If you we don’t consider the systematic velocity into consideration, then the maximum residual velocity is just simply the mean of the two components as we read them from the residual datacube, meaning [90 + 160 ]/2 km/s =125 km/s

For NGC 4845, the model reproduces the observation reasonably well in the kinematic major axis, leaving almost no residuals. However, the model leaves small residuals in the nuclear region and either sides of the kinematic minor axis (see Fig. \ref{Barolo-4845}).
The residuals reach a maximum velocity of $\sim 60\, \rm{km\,s^{-1}}$, that is blueshifted to the north-east of the major axis, south and south-east of the kinematic minor axis %with velocity $\sim 60\, \rm{km\,s^{-1}}$
and redshifted to the north and north-west of the minor kinematic axis with velocity $\sim$50 km s$^{-1}$. %The blueshifted velocity of NGC 4845 is smaller than the systemic velocity, and we consider the redshifted velocity as a maximum velocity of the gas.
%The residual datacube shows $\sim -60\rm{km\,s^{-1}}$ (blue shifted) and $\sim 50\rm{km\,s^{-1}}$ (redshifted ).  %The systematic velocity of this galaxy is ~ -133.4 km/s (as best estimated by BAROLO).  The systematic velocity is greater than the blue shifted component,  meaning the indicated blue shifted component is due to the systematic velocity of the galaxy rather than the gas motion, and hence I did not consider this component of the gas in the maximum residual velocity calculation. The maximum residual velocity is just the redshifted component, which is 50 km/s.
%Note that: If you we don’t consider the systematic velocity into consideration, then
The maximum residual velocity turns out to be in this case %is just simply the mean of the two components as we read them from the residual datacube, meaning [60 + 50 ]/2 km/s =
$\sim 55\, \rm{km\,s^{-1}}$.

%Note that: If you we don’t consider the systematic velocity into consideration, then the maximum residual velocity is just simply the mean of the two components as we read them from the residual datacube, meaning [60 + 50 ]/2 km/s = 55 km/s

For MCG-06-30-15, the model reproduces well the observed CO(2-1) kinematics in the nuclear region and along both the kinematic minor and major axes (see Fig. \ref{Barolo-mcg}).\\
The residuals reach a maximum velocity of $\sim 20\, \rm{km\,s^{-1}}$, that is blueshifted to the east direction with velocity $\sim$51 km s$^{-1}$ and redshifted to the south and south-west of the minor axis with velocity $\sim 20\, \rm{km\,s^{-1}}$. %The systemic velocity of MCG-06-30-15 is = $31.416 \, \rm{km\,s^{-1}}$, and therefore, the redshifted velocity of the gas is calculated taking the systemic velocity into consideration. 
%The residual datacube shows ~ – 20 km/s (blue shifted) and ~ 51 km/s (redshifted ).  %The systemic velocity of this galaxy is ~ 31.416 km/s (as best estimated by BAROLO). The systemic velocity of this galaxy is positive,  so it has to be compared with the redshifted component of the gas. The true redshifted component is [51-31.416] km/s  ~20 km/s. Therefore, the maximum residual velocity is now [20 + 20 ]/2 km/s = 20 km/s.
The maximum residual velocity is $\sim 35\, \rm{km\,s^{-1}}$.

%Note that: If you we don’t consider the systematic velocity into consideration, then the maximum residual velocity is just simply the mean of the two components as we read them from the residual datacube, meaning [20 + 51 ]/2 km/s = 35.5 km/s

Comparing these residual velocities of the gas in each galaxy  with the velocities the regular rotation shows us how important is the residual velocity (which could be due to devation from circular motions) for each galaxy. For NGC4968, the gas rotates at $\sim 200\, \rm{km\,s^{-1}}$ and there is a residual of $\sim 125\, \rm{km\,s^{-1}}$, which is a large fraction of the rotation velocity.\\
In NGC4845, the rotational velocity of the gas is $\sim 200\, \rm{km\,s^{-1}}$ (taking the average of the two components using the input ALMA datacube), and a residual with velocity about $\sim 55\, \rm{km\,s^{-1}}$ is detected. In MCG-06-30-15, the CO rotation velocity is $\sim 200\, \rm{km\,s^{-1}}$ and the residual velocity $\sim 35\, \rm{km\,s^{-1}}$  which is small with respect to the regular rotation velocity. 

For all galaxies the residual velocities are smaller than the main rotation velocity, indicating that the non circular kinematics very unlikely are tracing powerful outflows (whatever the cause is stellar or AGN related). Of all, the residual velocity of NGC4968 is larger than the other two galaxies with respect the rotation velocity. The least residual velocity is found for MCG, indicating that the kinematics of the gas in the central 1 kpc scale of this galaxy is best described by circular motions.

\subsection{Residuals from DiskFit}\label{resD}
As shown by the kinematic maps in Figs. \ref{Diskfit-4968}, \ref{Diskfit-4845}, and \ref{Diskfit-mcg} for NGC 4968, NGC 4845, and MCG-06-30-15, respectively, different DiskFit models reproduce the kinematics of the molecular gas as outlined by the CO(2-1) emission. Bisymmetric DiskFit models reproduce the observation much better than pure rotation DiskFit models for NGC 4968 and we consider this model as the best fit.
The $\chi^2$ reduces by $\sim$50\% for the radial motion model and by $\sim$63\% for bisymmetric model with respect to its value for pure rotation (flat disk) model, 
indicating that bisymmetric model is the best physical model for the kinematics of the gas in the central region of NGC 4968.
This leads us to conclude that the residuals seen in the pure rotation models (see Fig. \ref{Diskfit-4968}) cannot be ascribed to additional velocity components not belonging to the rotational kinematics. 
However, still there are residuals left over from the best-fit model, mainly in the south direction of the minor kinematic axis and west of the major axis (see Fig. \ref{Diskfit-4968}).
Also the model rotation plus radial motion DiskFit reduces the residual observed in the pure rotation DiskFit model (see rotation plus radial motion DiskFit model in Fig. \ref{Diskfit-4968},  middle panels), and indicates the presence of radial motions in the nuclear region of this galaxy.

Similarly, for NGC 4845, pure rotation model reproduces the observation well (see upper panels in Fig. \ref{Diskfit-4845}), however, bisymmetric model performes slightly better, leaving comparably smaller residuals in the kinematic minor axis (see lower panels in Fig. \ref{Diskfit-4845}).
In this case the change in $\chi^2$  is small when bisymmetric flows are included, reducing $\chi ^{2}$ by $\sim$30\%. We consider, however, that the bisymmetric
 model is the best-fit model.

Pure rotation (flat disk) DiskFit model reproduces the observation well for MCG-06-30-15, leaving only very small residuals (see upper panels in Fig. \ref{Diskfit-mcg}), and this can be considered the best-fit model for this galaxy. For MCG-06-30-15, unlike NGC 4968, incorporating a radial component into a pure rotation DiskFit model (rotation plus radial motion model) does not reduce the residuals observed in the pure rotation model, indicating the absence of significant radial motions in the nuclear region of this galaxy. 

%\subsubsection{Comparison with optical data}

%The HST images (left panels) and CO intensity map (right panels) of each galaxy are given in Fig. \ref{HST-CO}. 
%The comparison of the HST images and CO intensity maps shows the absence of CO emission in the very centre of NGC4968 (see table~\ref{t1}) while CO(2-1) emission does come from the centre too in the other two objects. Together with the kinematical analysis discussed above we can argue that this result can be linked to the presence of the bar in the inner part (see Fig.~\ref{NGC4968torque}). Indeed, the torque is positive inside (meaning the gas moves outward) and negative outside (the gas flows inward) this could be the reason why there is no CO in the centre of NGC4968.

\section{Line luminosity and molecular gas mass estimates}\label{MH2}

\begin{table}
\caption{\label{t3} Results of a Gaussian fit to the total CO(2-1) spectrum. \textit{Upper panel}: best-fit parameters of the Gaussian model components shown in Fig.~\ref{CO-Line}: central velocity ($\mu_{v}$)[km s$^{-1}$], flux amplitude (S$_{peak}$) [mJy], and velocity dispersion ($\sigma_{v}$) [km s$^{-1}$]. The number in the parenthesis indicates the Gaussian components. \textit{Lower panel}: Area over which the line emission is integrated on the collapsed image, the total CO(2-1)) line flux (S$_{CO}$) [Jy km s$^{-1}$], total CO(2-1) line luminosity ($L^\prime_{CO}$) [K km s$^{-1}$ pc$^{2}$] and molecular gas mass (M(${\rm H_2}$) [M$_{\odot}$]. Here we report the molecular mass values corresponding to $\alpha$=0.8. }
\centering
\setlength{\tabcolsep}{4pt}
\begin{tabular}{lllll}%{llccr}
\hline\\
Parameters&NGC 4968&NGC 4845&MCG-6-30-15\\
\hline\\
$\mu_{\rm v}$(1)&-11$\pm$3&-21$\pm$4& 125$\pm$3&\\
$\mu_{\rm v}$(2)& &-238$\pm$4&-8$\pm$8&\\
$S_{\rm peak}$(1)&0.71$\pm$0.01 &4.85$\pm$0.22&0.26$\pm$0.03&\\
$S_{\rm peak}$(2)& &5.38$\pm$0.22&0.22$\pm$0.01&\\
$\sigma_{\rm v}$(1)&258$\pm$6&157$\pm$10&67$\pm$9&\\
$\sigma_{\rm v}$(2)& &146$\pm$9&171$\pm$21&\\
\hline\\
Area & $3^{\prime\prime} \times 1.4^{\prime\prime}$ & $13^{\prime\prime} \times 3^{\prime\prime}$ & $3.4^{\prime\prime} \times 1.2^{\prime\prime}$\\
S$_{\rm CO(2-1)}$& 36& 617 & 16&\\
L$^{\prime}_{\rm CO(2-1)}$&$4 \times 10^{7}$&$12 \times 10^{7}$&$ 1.3\times 10^{7}$&\\
M$_{\rm tot}({\rm H_2})$  &$ 3 \times 10^{7}$&$ 9 \times 10^{7}$&$ 1 \times 10^{7}$&\\
M$_{\rm res}({\rm H_2})$ &$ 1 \times 10^{7}$&$0.3 \times 10^{7}$ &$ 0.07\times 10^7$&\\
\hline\\
\end{tabular}\\ \vspace{.05cm}
\end{table}

\begin{figure}
   \centering
   \includegraphics[width=9cm,height=5cm]{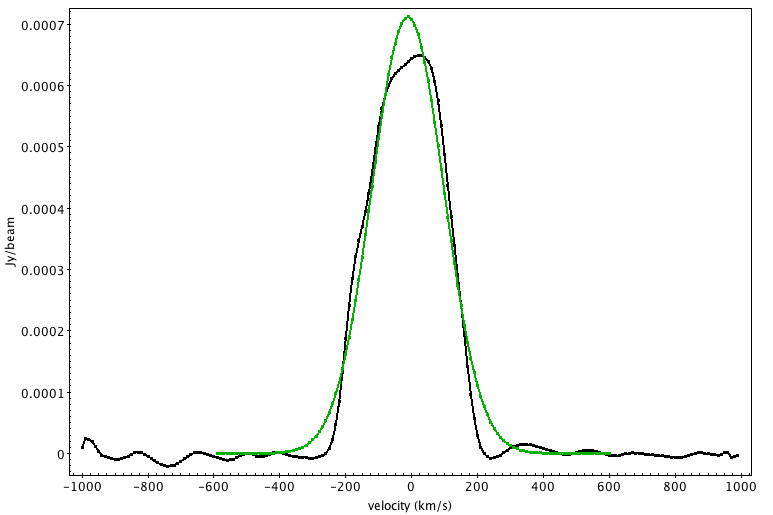}
   \includegraphics[width=9cm,height=5cm]{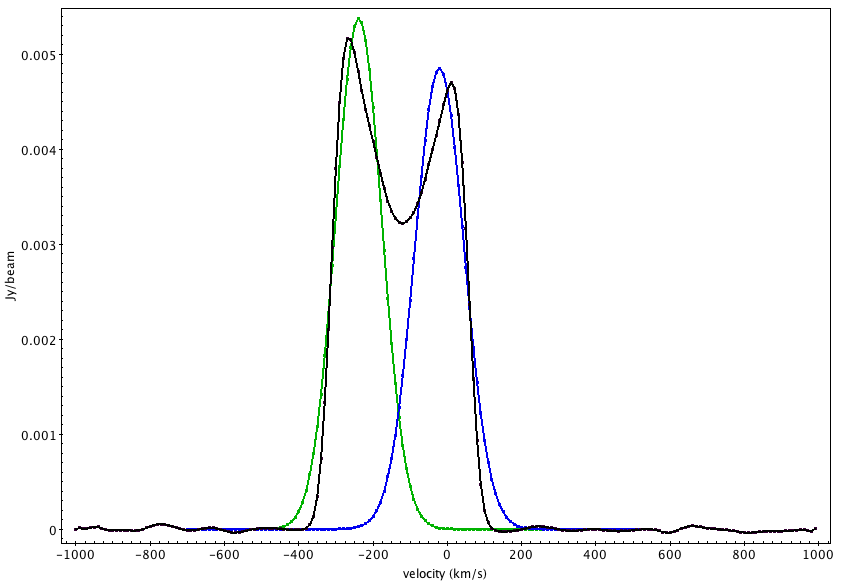}
   \includegraphics[width=9cm,height=5cm]{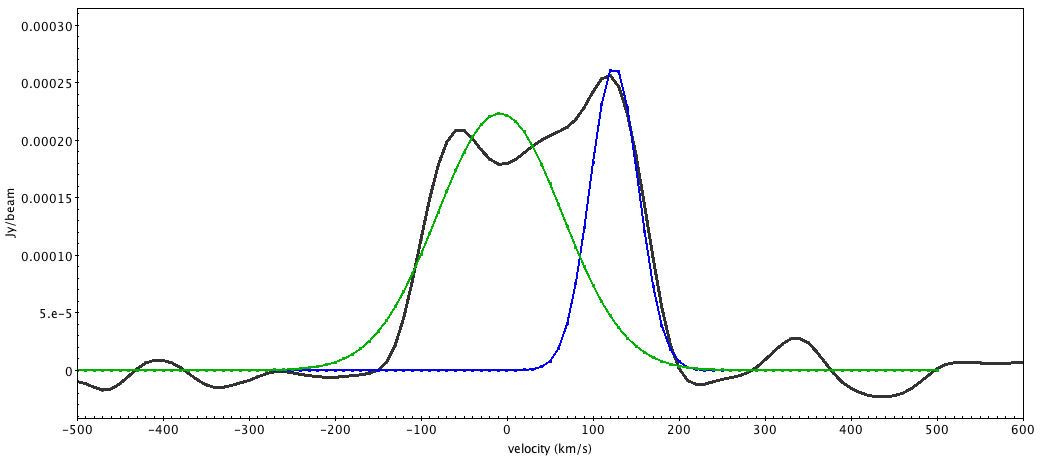}
   \caption{The Gaussian fit for the profile line of NGC 4968 (top panel), of NGC 4845 (middle panel), and of MCG-06-30-15 (bottom panel). Data are shown by the black line, while the fitted components by the blue and green lines (see detail in Table~\ref{t3}).}
   \label{CO-Line} 
   \end{figure}

To estimate the line luminosity ${\rm L^{\prime}_{CO}}$ and the molecular gas mass in the disc M(H$_{2}$),
we present the profile of the integrated CO(2-1) emission line in Fig. \ref{CO-Line}. The line profiles are generally smooth enough to be well fitted with a single (NGC 4968) or two (NGC 4845 and MCG-06-30-15) Gaussian profiles. We integrate the line emission over the moment0 maps. The results of the best fit parameters are given in Table \ref{t3} together with the size of the region over which the line flux is estimated. This was chosen by selecting those pixels where the S/N is larger than 3, roughly corresponding to 10\% of the maximum values. Then we derive the emission line luminosity ${\rm L^{\prime}_{CO}}$ using the relation given by \citet{solomon2005molecular}:

   \begin{equation}\label{eq1}
     L^{\prime}_{CO} = 3.25 \times 10^{7} S_{CO}\Delta \varv v_{obs}^{-2}D_{L}^{2}(1+z)^{-3},
     \end{equation}
where $L^{\prime}_{CO}$ is the CO line luminosity given in K km s$^{-1}$ pc$^{2}$, $S_{CO}\Delta \varv$ is the velocity-integrated flux in Jy km s$^{-1}$, $\nu_{obs}$ is the observed frequency in GHz; $z$ is the redshift, and $D_L$ is the luminosity distance in Mpc ($D_{L} = D_{A}(1+z)^{2}$, $D_{A}$ is the angular size distance) is the luminosity distance in Mpc. The molecular mass, ${\rm M(H_2)}$ is estimated using the relation (\cite{solomon2005molecular}):
 \begin{equation}\label{eq2}
  M(H_{2}) = \alpha_{CO} L^{\prime}_{CO},
  \end{equation}
where $\alpha_{\rm CO}$ is the CO-to-H$_{2}$ conversion factor. The value of $\alpha_{\rm CO}$ is highly uncertain, it depends on metallicity and environment \citep[see i.e.][]{bol+13} and may vary in the range 0.8-3.2. For simplicity we use here the lower value which is the average value in active galaxies.
Furthermore, strictly speaking the conversion factor entails the CO(1-0) line luminosity, this means that in absence of similar CO(1-0) observations an assumption on the excitation status of the molecular gas must be made. We use the average CO spectral line distribution of \citet{kam+16} corresponding to the $\log ({\rm L_{FIR}})$ luminosity range of 10-10.5~${\rm \log L_\odot}$ to convert from the observed ${\rm L^{\prime}_{CO(2-1)}}$ to ${\rm L^{\prime}_{CO(1-0)}}$ which is used in eq.~\ref{eq1}. The resulting ratio is $\frac{{\rm L^{\prime}_{CO(2-1)}}}{{\rm L^{\prime}_{CO(1-0)}}}=2/3$. For NGC4968 we use the ratio measured by \citet{strong2004molecular}

The total gass mass M$_{\rm tot}({\rm H_2})$ in the disc of the three galaxies is listed in Table \ref{t3}, which agrees well with the typical values for other nearby and low-luminosity AGNs, and are in line with the claim that Seyfert 2 galaxies do seem to possess more molecular mass than Seyfert 1 galaxies (see \citet{strong2004molecular} and references therein)
and agree with the single dish values published by \citet{strong2004molecular} for NGC4968 and MCG-06-30-15 and by \citet{Rosario} for MCG-06-30-15.\\
Similarly, we determine the line luminosity and molecular gas mass in the modelled disc of each galaxy. The molecular gas mass fraction in the modelled disc of NGC 4968, NGC 4845 and MCG-06-30-15 corresponds to $\sim$69\%, $\sim$97\%  and $\sim$92\% of the total molecular gas mass in each galaxy, respectively. 

In addition to the total molecular gas mass and the gas mass in the main rotating disc, we also estimated the line luminosity and then the molecular mass of the residuals for each galaxy using Eqs. \ref{eq1} and \ref{eq2} and  listed in Table~\ref{t3}.
The residuals are the emissions resulting from the substraction of the modelled from the observed data cube. %The CASA task {\it imfit} is used to obtain the integrated CO(2-1) flux. 

\section{Gas and continuum distribution}\label{dust}

\begin{figure}
   \centering
   \includegraphics[width=0.48\textwidth,angle=0]{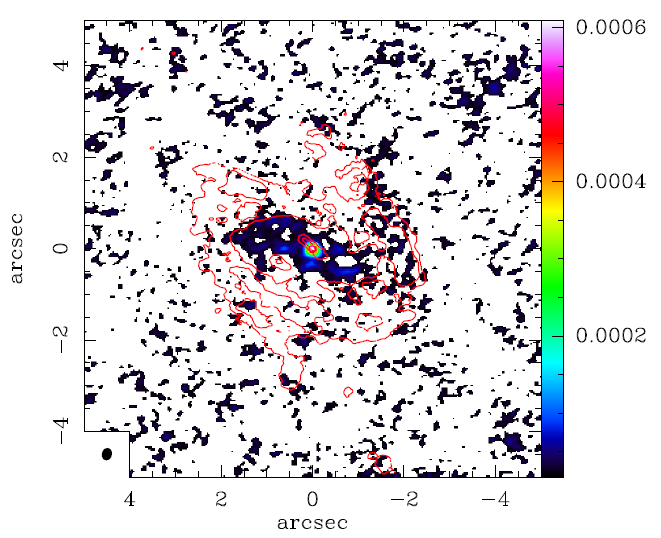}
   \includegraphics[width=0.48\textwidth,angle=0]{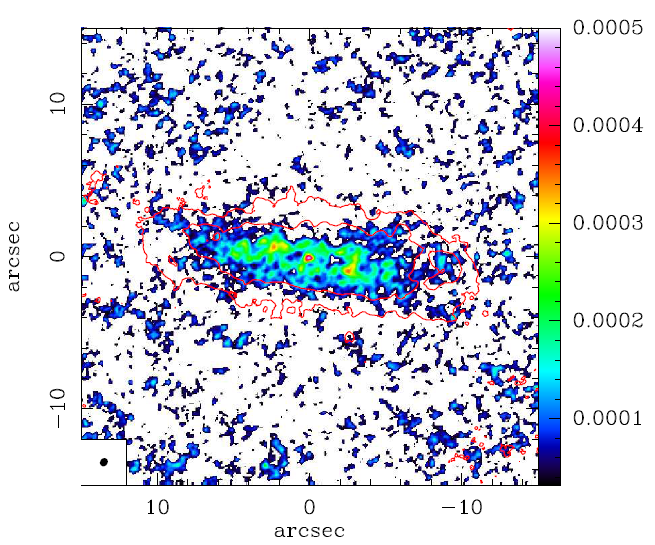}
   \includegraphics[width=0.48\textwidth,angle=0]{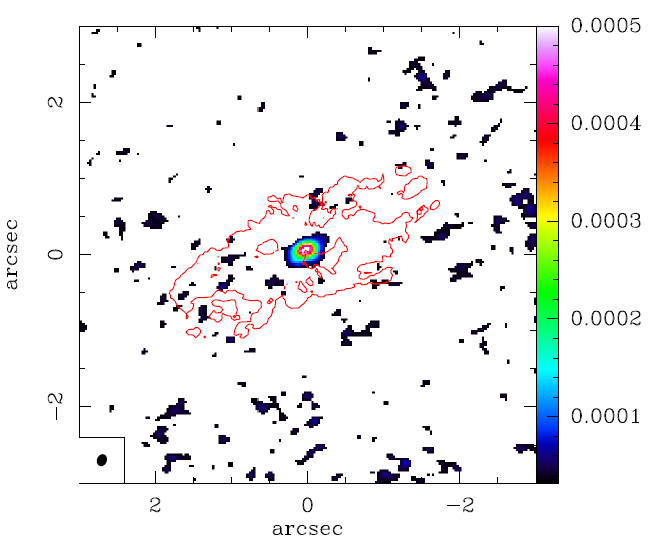}
   \caption{Overlay of CO(2-1) contours on the 1.2 mm continuum image of NGC 4968 with 2 CO contours of 4 and 16\% of maximum  (top panel), of NGC 4845 with 2 CO contours of 5 and 20\% of maximum (middle panel), and of MCG-06-30-15 with 2 CO contours of 5 and 20\% of maximum (bottom panel). The continuum scale (colour wedge) is in Jy/beam. The 3$\sigma$ level is 0.1 mJy, therefore 0.0001 Jy/beam
in the units of the Figure for N4968, 0.5 mJ=y/beam, then 0.0005 Jy/beam for N4845 and 0.05 mJy/beam therefore 0.00005 for MCG-06-30-15.}
      \label{CO-dust}        
    \end{figure}

Figure~\ref{CO-dust} shows the overlay of the collapsed CO intensity map with the map of the continuum at the observing frequencies, derived from the flux averaged over
all the four spectral windows once neglecting the channels with the line emission.
 CO emission is absent from the centre of NGC4968 and not in the other two galaxies. Together with the kinematical analysis discussed above we can argue that this result can be linked to the presence of the bar in the inner part (see Fig.~\ref{NGC4968torque}). Indeed, the torque is positive inside (meaning the gas moves outward) and negative outside (the gas flows inward) this could be the reason why there is no CO in the centre of NGC4968.

The continuum distribution is overall more compact than the CO emission especially in NGC4968 and MCG-06-30-15. One may argue that some flux could be lost because of the high resolution of these observations and the lack of additional compact array observations. However because of the integrated CO flux agrees with that measured with single dish observations \citep{strong2004molecular,Rosario}, at least in NGC4968 and MCG-06-30-15, the amount of missing flux should not be significant. 

The origin of the majority of the continuum emission is very likely due to dust. The central nucleus is expected to emit synchrotron emission from the black hole jet or corona which may have a significant contribution to the radiation in the millimetre range. However,  the three galaxies are strong far-IR emitters as detected by infrared satellites \citep[see i.e.][]{grup+16,cor+14,mel+14} and have a radio spectrum typical of a central synchrotron source with a spectral index decreasing with frequencies as $\nu\sim -0.6\div -0.8$ (radio data taken from \citet{con,mun}), which contributes a few percent at 1\,mm wavelength.

%\begin{figure}
 %  \centering
 %  \includegraphics[width=0.48\textwidth,angle=0]{n4968-comp.PNG}
%   \includegraphics[width=0.48\textwidth,angle=0]{n4845-comp.PNG}
%   \includegraphics[width=0.48\textwidth,angle=0]{mcg-comp.PNG}
%   \caption{\textit{Top panels}: The F606W HST image (top left) and the CO intensity map (top right) of NGC 4968. \textit{Middle panels}: The F814W HST  and CO intensity images for NGC 4845. \textit{Bottom panels}: The F160W HST and CO intensity images for MCG-06-30-15. The color bars are in mJy/beam kms$^{-1}$ for CO maps.}
%      \label{HST-CO}        
%    \end{figure}
    %\clearpage

\section{Discussion}\label{dis}

\subsection{Kinematical perturbations}\label{kinpert}
In addition to the kinematic maps, the comparison of the $p-v$ diagrammes of the host galaxy disc (blue contours) with the fitted $^{3D}$BAROLO model of a rotating disc (red contours) along the kinematic major and minor axes (see bottom panels in Figs. \ref{Barolo-4968}, \ref{Barolo-4845}, and \ref{Barolo-mcg}) further shows how well the model fits the data, and helps to see the presence of any deviation from circular motions. For NGC 4968, along the kinematic major axis, the $^{3D}$BAROLO model with the rotating disc (red contours) fits the rotation curve of the host galaxy disc (blue contours) relatively well, leaving only small room for additional significant kinematical components (see the major axis in the bottom panel in Fig. \ref{Barolo-4968}).
The $^{3D}$BAROLO rotating disc model fits also well the observation in the nuclear region and along the kinematic minor axis of the host galaxy disc, however, leaving an important deviation at the end of the kinematic minor axis (see the minor axis in the bottom panel in Fig. \ref{Barolo-4968}), which is in agreement with what is observed in the corresponding velocity map in the same figure.

In NGC 4845, the $^{3D}$BAROLO model with the rotating disc fits the rotation curve of the host galaxy disc along the kinematic major axis (see the $p-v$ diagram in Fig. \ref{Barolo-4845}). Along the kinematic minor axis the $^{3D}$BAROLO model also fits relative well the observations. However it fails {\it around} the kinematic minor axis we see small deviations (see $p-v$ diagram in Fig. \ref{Barolo-4845}).

In MCG-06-30-15, the rotating disc model fits the rotation curve of the host galaxy disc quite well both along the kinematic major and minor axes, presenting only very small deviations (see the $p-v$ diagram in Fig. \ref{Barolo-mcg}), further strengthening what is observed in the corresponding kinematic map (Fig. \ref{Barolo-mcg}). 

Deviations from circular motions could be caused by different mechanisms, such as outflows driven by AGN or star formation. Also, it has been known that non-circular motions of molecular gas in the nuclear regions of disc galaxies could be due to the existence of a bar like structure, a warped circumnuclear disc, or radial motion due to other mechanisms. For example, using lower resolution CO(2-1) observations, \cite{Schinnerer2000} observed non-circular motions caused by warped structure in the nuclear region (approximately radial distances of 0.7-1$^{"}$) of NGC 3227. Furthermore, it has been shown that gas inflows could be due to gravity torques from non-axisymmetric potentials in the central regions of galaxies, such as streaming motions along a bar (see e.g., \citealt{ruffa2019agn}).

Indeed, the presence of residuals/deviations could be either due to some gas components not included within the main rotating disc considered by the model (this could be due to difference in geometry with the main rotating disc), or there is a deviation in the kinematics of the gas from circular motion, or both. In the galaxies of this study the observed residuals could be an indication for the presence of deviations in the kinematics of the gas from circular motions, as shown by the kinematic maps and corresponding $p-v$ diagrams. 

Bisymmetric Diskfit model performes better in reproducing the observations than other DiskFit models in NGC 4845 and NGC 4968, indicating that the kinematical perturbations in these two galaxies are more likely due to a bar pattern, whereas pure rotation model is the best-fit model for MCG-06-30-15. Note that the bisymmetric DiskFit model describes an elliptical or a bar like flow, and it is not surprising that bisymmetric model appears to be the best-fit model in NGC4968 and NGC 4845, since the galaxies are shown to be barred (see more below).

As revealed by both models, we argue that although the circular motion is the dominant kinematics in the molecular disc of all galaxies, there is clear evidence for the presence of non-circular motions in the nuclear regions of NGC 4968 and NGC 4845, mainly in NGC 4968, where non-circular motions appeared to be significant (see Sections~\ref{resB} and~\ref{resD}). However, the smallness in the width of the residual velocity (compared to the circular velocity) indicates the absence of energetic feedback both from the central AGN and star formation in the nuclear regions of the galaxies. Also, the star formation rate in all galaxies is very small (see Table \ref{t1}).

Table \ref{t1} lists too the morphological types from either NED or Hyperleda: N4968 has a strong bar and the torque analysis (see below) also confirms that.
We show in Fig. \ref{NGC4968rotation} a possible model for the rotation curve in the central kpc of the galaxy, built from its various components, bulge, disc of stars and gas, black hole and dark matter. The latter components contribute negligibly, the dark matter is required however, to explain a flat rotation curve in the outer parts. The bulge mass was taken from the decomposition of the red image, used in the potential calculation, and calibrated with the rotation curve. The black hole mass has been assumed to be 7$\times10^6$ M$_\odot$, about 0.2\% of the mass of the bulge, according to scaling relations (e.g. \citealt{kormendy2013coevolution}). The bulge dominates the region of interest, for the observed molecular component. Thus the mass decomposition allows the computation of the precessing frequency $\Omega-\kappa/2$, and the determination of Lindblad resonances. If we place the corotation just outside the bar, as it is frequently observed in barred spirals (e.g. \cite{1996FCPh...17...95B}), then the pattern speed of the bar is $\Omega_b$ = $52\, \rm{km\,s^{-1}\,kpc^{-1}}$, putting corotation at 3.5 kpc and inner Lindblad resonance (ILR) at R = 300pc, corresponding to the CO ring. \\

N4845 morphology has a nice peanut shape bulge, and this is well known to be due to a bar (e.g., \citealt{1990A&A...233...82C}). From our analysis there is evidence of an additional weak kinematics but we cannot argue that this is due to the presence of the bar. Our kinematical analysis seems to confirm that unlike the one in NGC 4968 the bar in NGC4845 is unable to change significantly the regular rotation pattern of the molecular gas. This additional weak kinematics (see Fig.~\ref{Barolo-4845} and~\ref{Diskfit-4845}) could be due to gas inflowing or outflowing the central region but with the present data there is not firm evidence of it.\\

\begin{figure}
  \centering
  \includegraphics[width=0.35\textwidth,angle=0]{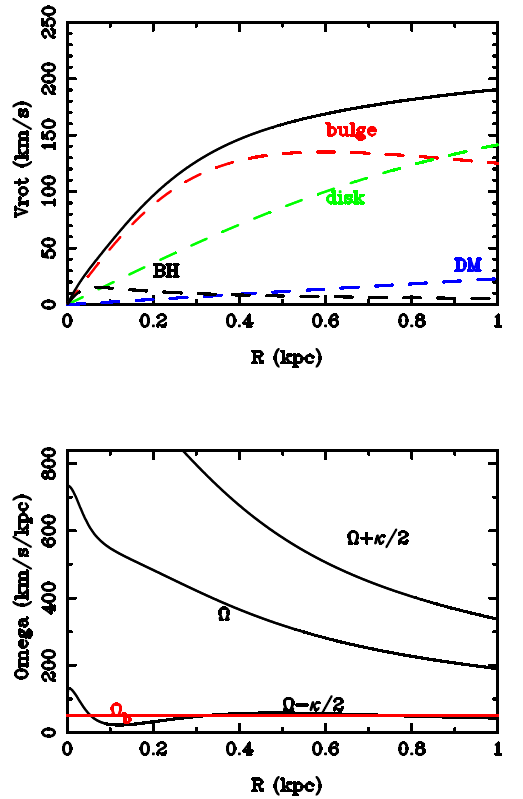}
 \caption{\textit{Top panel}: A possible model for the rotation curve in the central kpc of NGC 4948, built from its various components, bulge, disc of stars and gas, black hole and dark matter. This latter contributes negligibly but it is required to explain a flat rotation curve in the outer parts (see text).
\textit{Bottom panel}: The precessing frequency of elongated orbits, $\Omega-\kappa/2$, used to model the orbits in the inner region of NGC 4968. The red line is the pattern speed of the bar (e.g. \cite{1996FCPh...17...95B}).}
 \label{NGC4968rotation}
\end{figure}

% TORQUE COMPUTATION FOR NGC 4968

\subsection{Torque computation for NGC 4968}\label{sec:torq}
To determine whether the gas is driven outwards or inwards, a calculation of the bar torques is necessary \citep[e.g.][]{audibert2019}. Through the Poisson equation, the gravitational potential is derived from the stellar density, assumed to be the main responsible to the gravity forces in the plane of the galaxy, inside the central kiloparsec. The dark matter is indeed negligible in the very centre (e.g. NGC 4968 in Fig.\ref{NGC4968rotation}). The stellar density is better traced by the HST H-band image (F160W), since in the NIR the dust impact is minimised (see \citet{comb+19} for details). The HST-NIC2 image reveals a central bulge and a stellar disc. To de-project the galaxy to face-on, we have first isolated the bulge, assumed to be spherical, which should not be de-projected. We then de-projected the galaxy disc with the adopted inclination angle of 60$^\circ$ and PA 250$^\circ$. The bulge was then added to the de-projected image. We assumed a constant mass-to-luminosity ratio and calibrated it to retrieve the observed rotation curve, modeled as shown in Fig. \ref{NGC4968rotation} (bottom panel). The gravitational potential is derived from the stellar distribution, assuming a thin disc of scale ratio $h_z/h_r$ =1/12. Both the HST-NIR and CO(2-1) de-projected images of the galaxy have been resampled to the same pixel size of 0.03 arcsec = 6~pc  \citep{comb+19}.

\begin{figure}
  \includegraphics[width=0.4\textwidth,angle=0]{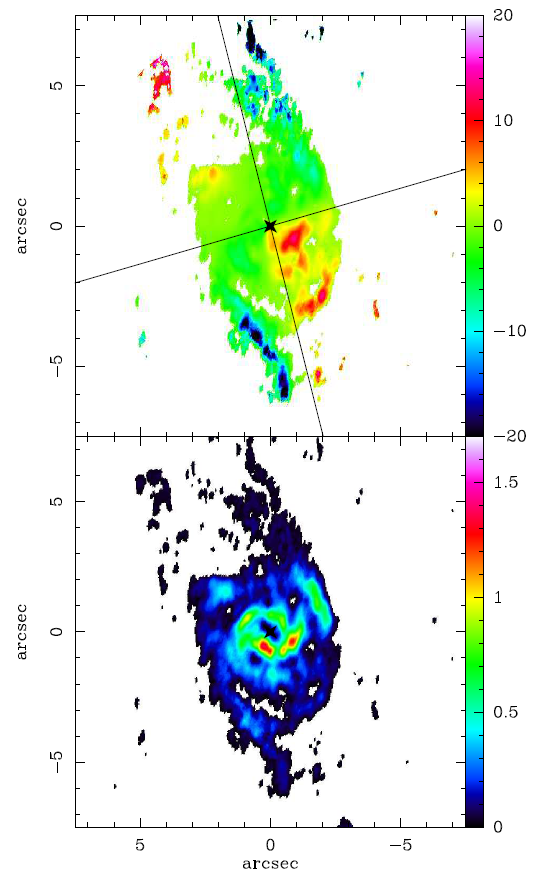}
\caption{{\it Top panel:} Map of the gravitational torques exerted on the gas  by the stellar potential, in the centre of NGC\,4968.  The map shows in each pixel the torque derived from the HST-NIR image, multiplied by the gas surface density. Both images have been de-projected to face-on (see text for details). The torques change sign as expected in a four-quadrant pattern (or butterfly diagram). The orientation of the quadrants follows the bar orientation. In this de-projected picture, the major axis of the galaxy is oriented parallel to the horizontal axis. {\it Bottom panel:} The de-projected image of the CO(2-1) emission, at the same scale, and with the same orientation, for comparison.}
\label{NGC4968torque}
\end{figure}

We have computed the bar gravity torques, following the method described in (e.g. \citealt{garcia2005torq}; \citealt{audibert2019}). The forces are computed at each pixel, by derivating the potential, and then the torques are computed on the gas, taking into account the gas density at the given pixel. The torque map is plotted in Fig. \ref{NGC4968torque} together with the de-projected gas surface density, for comparison. As shown in the top panel of the figure, the torque map reveals the expected butterfly diagram (four-quadrant pattern) in relation to the bar orientation (indicated by straight lines) with torques changing sign in each quadrant. This implies that the observed non-circular motions are well fitted by the gas flow in a bar potential.

    %\clearpage

\begin{figure}
 \includegraphics[width=0.4\textwidth,angle=0]{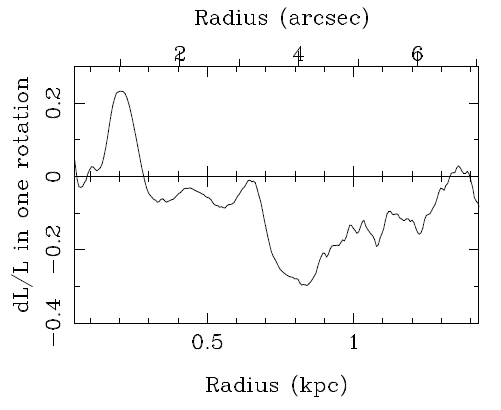}
\caption{The radial distribution of the average torque exerted by the stellar bar on the gas of NGC 4968. The torque is normalised to show the fraction of the angular momentum transferred from the gas in one rotation--$dL/L$, estimated from the CO(2-1) de-projected map. The curve is plotted only from a radius = 0.23 arcsec (or 46~pc), which is the largest size of the beam.}
\label{NGC4968avetorque}
\end{figure}

 From the torque map we compute azimuthal averages and obtain the effective torque at each radius, per unit mass. This yields the derivative of the gas angular momentum at this radius. To derive the relative variation of angular momentum, we divide by the average angular momentum at this radius (from the rotation curve). This relative variation of angular momentum in one rotation is plotted in Fig. \ref{NGC4968avetorque}. The torque is positive in the very centre, meaning that the gas is driven outwards to the gas ring. Outside of the ring, the torque is negative and peaks at dL/L = -0.3 in one rotation: the gas is driven inwards, to accumulate in the ring. In this region the gas is losing about 30\% of its angular momentum in one rotation.\\
The NLR dynamics does not uniquely show the presence of an outflow because it might be affected by extinction \citep{ferr2000}. If an ionised outflow were present, at least in the central region,  the outward motions of the gas (as revealed by the positive torque in the very centre) could be partly interpreted as the interaction between the ionized outflows and the gas. However, since the torque is negative in the outside of the ring, the outflows could be counter balanced by inward motions.

%\clearpage
\section{Summary and Conclusions}\label{con}
We have analysed the properties and kinematics of the cold molecular gas in the nuclear and circumnuclear regions of three Seyfert galaxies, NGC 4968, NGC 4845 and MCG-06-30-15, using ALMA observations of their CO(2-1) emission line. We used the $^{3D}$BAROLO and DiskFit (both axisymmetric and non-axisymmetric) softwares to model the kinematics of the molecular gas.\\
The main findings are summarised as follows: 
\begin{itemize}
\item The intensity maps reveale a ring-like morphology in the nuclear regions of NGC 4968 and MCG-06-30-15, whereas the shape is edge-on in NGC 4845.\\ 
%\item The comparison of the HST images with the CO intensity maps shows the absence of CO in the centre of NGC 4968 contrary to MGC-06-30-15, and the dust distribution roughly traces the CO in and around the centre of NGC 4845.\\
\item The gas kinematics in the molecular discs of NGC 4845 and MCG-06-30-15 is dominated by pure rotational motion, but there is evidence for non circular-motions in NGC 4968
and weakly in NGC 4845.\\
\item Unlike NGC 4845, where the deviation from circular motion is small, a significant non circular motion are observed in NGC 4968, mainly along the kinematic minor axis, with velocity $\sim 115\, \rm{km\,s^{-1}}$, which are likely due to the bar.\\
\item Moreover, of all DiskFit models, the bisymmetric  model is found to be the best-fit model for NGC 4968 and NGC 4845, in agreement with the nuclear bar origin. \\

\item Regular rotation is shown to be the dominant kinematics of the gas in the nuclear region of MCG-06-3-15, and hence pure rotation model is the best fit model for this galaxy.

\item The molecular mass, ${\rm M(H_{2}}$), in the nuclear disc is estimated to be
 $\sim 3-12\times 10^{7} ~{\rm M_\odot}$ (NGC 4968), $\sim 9-36\times 10^{7}~ {\rm M_\odot}$ (NGC 4845), and $\sim 1-4\times 10^{7}~ {\rm M_\odot}$ (MCG-06-30-15), allowing the CO-to-H$_{2}$ conversion factor $\alpha_{CO}$ between 0.8 and 3.2, typical of nearby galaxies of the same type.
\item The molecular gas mass of the modeled disc of each galaxy corresponds to $\sim$ 69\%, $\sim$97\% and $\sim$92\% of the total molecular gas mass in NGC 4968, NGC 4845 and MCG-06-30-15, respectively.\\
\item For the galaxy NGC~4968, placing the corotation just outside of the bar indicates a bar pattern speed of  $\Omega_b$ = $52\, \rm{km\,s^{-1}\,kpc^{-1}}$, putting corotation at 3.5 kpc and inner Lindblad resonance (ILR) ring at R = 300pc, which corresponds to the CO ring.\\
The computation of the torques exerted by the stellar bar on the gas shows that the torques are positive inside the molecular ring and negative outside, revealing that the gas is accumulating in the inner Lindblad resonance. Thus, the observed non-circular motions in the molecular disc of  NGC 4968, could be due to the presence of the bar in the nuclear region.

\end{itemize}

In summary, in the studied galaxies the radiative feedback or winds, seen in the kinematics of emission lines from ionised gas (see \S~\ref{pppt}), 
do not substantially alter the environments within molecular clouds that produce the bulk of the low-excitation CO emission.
We do not find any strong evidence for gas kinematics close to the central AGN which might be univocally attributed to the effect of the AGN. The cold, star-forming molecular gas in the centre of the host galaxy is not strongly influenced by the presence of the AGN, despite the fact that these AGN are luminous enough to dynamically disturb this material.\\
We cannot claim any strong evidence in these sources of the long sought feedback/feeding effect due to the coupling of mechanical energy from the nucleus with the cold star-forming phase.

\section*{Acknowledgements}
This paper makes use of the following ALMA data: ADS/JAO.ALMA$\#$2017.1.00236.S.ALMA is a partnership of ESO (representing its member states), NSF (USA) and NINS (Japan), together with NRC (Canada), MOST and ASIAA (Taiwan), and KASI (Republic of Korea), in cooperation with the Republic of Chile. The Joint ALMA Observatory is operated by ESO, AUI/NRAO and NAOJ. The National Radio Astronomy Observatory is a facility of the National Science Foundation operated under cooperative agreement by Associated Universities, Inc. This publication has made use of data products from the NASA/IPAC Extragalactic Database (NED). We acknowledge the usage of the HyperLeda database (http://leda.univ-lyon1.fr). A. Bewketu Belete acknowledges the support by the Brazilian National Council for Scientific and Technological Development (CNPq). R.S. acknowledges the support provided by CONICYT through FONDECYT postdoctoral research grant No. 3200909. JAFO and LS acknowledge financial support by the Agenzia Spaziale Italiana (ASI) under the research contract 2018-31-HH.0.
Research activities of the Observational Astronomy Board of the Federal University of Rio Grande do Norte (UFRN) are supported by continuous grants from CNPq, CAPES and FAPERN brazilian agencies. J.R.M. and A.B.B also acknowledge financial support from INCT INEspaço/CNPq/MCT.
An anonymous referee is warmly thanked for their comments/suggestions have greatly improved the completeness of the paper.
%\clearpage


\begin{thebibliography}{t!}
\bibitem[Alonso-Herrero et al. (2018)]{2018ApJ...859..144A} {Alonso-Herrero}, A., {Pereira-Santaella}, M.,  {Garc{\'\i}a-Burillo}, S. et al. 2018, ApJ, 859, 144
\bibitem[Alonso-Herrero et al. (2019)]{2019A&A...628A..65A} {Alonso-Herrero}, A., {Garc{\'\i}a-Burillo}, S., {Pereira-Santaella}, M., {Davies}, R.~I., {Combes}, F., 2019, A\&A, 628, 65
\bibitem[Audibert et al. (2019)]{audibert2019} Audibert, A., Combes, F., Garcia-Burillo, S. et al.: 2019, A\&A 632, A33
\bibitem[Bertola et al. (1989)]{bertola1989evidence} Bertola, F., Rubin, V.C., Zeilinger, W. W, 1989, ApJ, 345, L29
\bibitem[Bolatto A., et al (2013)]{bol+13} Bolatto A., et al., 2013 ARA\&A 51, 207
\bibitem[Brighjtman \& Nanda (2011)]{nanda} Brightman M., Nanda K.,  2011, MNRAS 413,1206B
\bibitem[Buta \& Combes (1996)]{1996FCPh...17...95B}{Buta}, R., {Combes}, F., 1996, Fundamentals of Cosmic Physics, 17, 95
\bibitem[Carniani et al. (2015)]{carniani2015ionised} Carniani, S., Marconi, A., Maiolino, R., Balmaverde, B., Brusa, M, 2015, A\&A, 580, 102
\bibitem[Cicone et al. (2014)]{cicone2014massive} Cicone, C.,  Maiolino, R., Sturm, E., Graci{\'a}-Carpio, J., Feruglio, C., 2014, A\&A, 562, A21
\bibitem[Combes et al. (1990)]{1990A&A...233...82C} {Combes}, F., {Debbasch}, F., {Friedli}, D., {Pfenniger}, D. et al. 1990, A\&A, 233, 82
\bibitem[Combes et al. (2013)]{2013A&A...558A.124C}{Combes}, F., {Garc{\'\i}a-Burillo}, S., {Casasola}, V. et al., 2013, A\&A, 558, A124
\bibitem[Combes et al. (2014)]{2014A&A...565A..97C} {Combes}, F., {Garc{\'\i}a-Burillo}, S., {Casasola}, V. et al., 2014, A\&A, 565, A97
\bibitem[Combes et al. (2019)]{comb+19} Combes, F., {Garc{\'\i}a-Burillo}, S., {Audibert}, A. et al., 2019, A\&A, 623, A79
\bibitem[Condon et al. (2002)]{con} Condon J. J., Cotton W. D., Broderick J. J., 2002, AJ 124, 675 
\bibitem[Cortese et al. (2014)]{cor+14} Cortese L. et al., 2014, MNRAS 440, 942
\bibitem[Dasyra \& Combes (2011)]{dasyra2011turbulent}Dasyra, K.M., Combes, F, 2011, A\&A, 533, L10
\bibitem[Dasyra \& Combes (2012)]{dasyra2012cold} Dasyra, K.M.,  Combes, F, 2012, A\&A, 541, L7
\bibitem[de Vaucouleurs (1991)]{Vaucouleurs1991} de Vaucouleurs, G., de Vaucouleurs, A., and Harold Jr, G. et al. 1991, Third Reference Catalogue of Bright Galaxies (New York: Springer) (RC3)
\bibitem[Di Teodoro \& Fraternali (2015)]{teodoro20153d} Di Teodoro, E.M.,  Fraternali, F. 2015, MNRAS, 451, 3021
\bibitem[Dom{\'\i}nguez-Fern{\'a}ndez et al. (2020)]{2020A&A...643A.127D} {Dom{\'\i}nguez-Fern{\'a}ndez}, A.~J., {Alonso-Herrero}, A., {Garc{\'\i}a-Burillo}, S. et al. 2020, A\&A, 643, A127 
\bibitem[{Fern{\'a}ndez-Ontiveros} et al. (2020)]{2020A&A...633A.127F} {Fern{\'a}ndez-Ontiveros}, J.~A., {Dasyra}, K.~M., {Hatziminaoglou}, E. et al.  2020, A\&A, 633, A127
\bibitem[Ferruit et al. (2000)]{ferr2000} Ferruit, P., Wilson, A. S., \& Mulchaey, J. et al. 2000, ApJS, 128, 139
\bibitem[Feruglio et al. (2010)]{feruglio2010quasar}Feruglio, C., Maiolino, R., Piconcelli, E., et al. 2010, A\&A, 518, L155
\bibitem[Fiore et al. (2017)]{2017A&A...601A.143F} {Fiore}, F., {Feruglio}, C.,  {Shankar}, F., {Bischetti}, M. {Bongiorno}, A., 2017, A\&A 601, A143
\bibitem[Fluetsch et al. (2019)]{fluetsch2019cold} Fluetsch, A., Maiolino, R., Carniani, S. et al. 2019, MNRAS, 483, 4586
\bibitem[Fukazawa et al. (2011)]{2011ApJ...727...19F} {Fukazawa}, Y., {Hiragi}, K., {Mizuno}, M. et al. 2011, \apj, 727, 19
\bibitem[Garc{\'\i}a-Burillo et al. (2005)]{garcia2005torq} García-Burillo, S., Combes, F., Schinnerer, E. et al.: 2005, A\&A 441, 1011
\bibitem[Garc{\'\i}a-Burillo et al. (2014)]{garcia2014molecular} Garc{\'\i}a-Burillo, S., Combes, F., Usero, A. et al. 2014, A\&A, 567, A125
\bibitem[Garc{\'\i}a-Burillo et al. (2015)]{garcia2015high} Garc{\'\i}a-Burillo, S., Combes, F.,  Usero, A. et al. 2015, A\&A, 580, A35
\bibitem[Garc{\'\i}a-Burillo et al. (2016)]{2016ApJ...823L..12G} Garc{\'\i}a-Burillo, S., Combes, F., Ramos Almeida, C. et al. 2016, A\&A, 823, L12 
\bibitem[Garc{\'\i}a-Burillo et al. (2019)]{2019A&A...632A..61G} Garc{\'\i}a-Burillo, S., Combes, F., Ramos Almeida, C. et al. 2016, A\&A, 632, A61
\bibitem[Gruppioni et al. (2016)]{grup+16} Gruppioni C., et al. (2016) MNRAS 458, 4320
\bibitem[Haan et al. (2009)]{Haan_2009} Haan, S., Schinnerer, E., Emsellem, E., et al. 2009, ApJ, 692, 1623
\bibitem[Irwin et al. (2015)]{Irwin_2015}  Irwin, J.A., Henriksen, R.N., Krause, M. et al. 2015, ApJ, 809, 172
\bibitem[Kamenentzky et al. (2016)]{kam+16} Kamenetzky J., Rangwala N., Glenn J., Maloney P.R. and Conley A., 2016, ApJ 829, 93 
\bibitem[Kanekar et al. (2008)]{kanekar2008outflowing}Kanekar, N., Chengalur, J. N, 2008, MNRAS 384, L6
\bibitem[Kormendy \& Ho (2013)]{kormendy2013coevolution}Kormendy, J., Ho, L. C, 2013, ARA\&A, 51, 511
\bibitem[LaMassa et al. (2017)]{lamassa2017chandra} LaMassa, S. M., Yaqoob, T., Levenson, N.A. et al.  2017, ApJ, 835, 91
\bibitem[Lutz et al. (2020)]{refId0} {Lutz, D.,} {Sturm, E.,} {Janssen, A.,} et al. 2020, A\&A, 633, A134
\bibitem[McMullin et al. (2007)]{2007ASPC..376..127M} {McMullin}, J.~P., {Waters}, B., {Schiebel}, D., {Young}, W. and {Golap}, K. et al. 2007, Astronomical Society of the Pacific Conference Series, 376, 127
\bibitem[Makarov et al. (2014)]{2014A&A...570A..13M} {Makarov}, D., {Prugniel}, P., {Terekhova}, N., {Courtois}, H., {Vauglin}, I., A\&A, 570, A13
\bibitem[Malkan et al. (1998)]{malkan1998hubble} Malkan, M. A., Gorjian, V., Tam, R., 1998, ApJ Sup., 117, 25
\bibitem[Marinucci et al. (2014)]{Marinucci_2014}Marinucci, A., Matt, G., Miniutti, G. et al. 2014, ApJ, 787, 83
\bibitem[Mel\'endez et al. (2014)]{mel+14} Mel\'endez M., Mushotzky R. F., Shimizu T. T., Barger A. J., Cowie L. L., 2014 ApJ 794, 152
\bibitem[Mordini et al. (2021)]{mor21} Mordini S., et al., 2021, in preparation
\bibitem[Morganti et al. (2005)]{morganti2005fast} {Morganti, R., Tadhunter, C. N., Oosterloo, T. A.}, 2005, A\&A, 444, L9
\bibitem[Morganti et al. (2015)]{2015A&A...580A...1M} {Morganti}, R., {Oosterloo}, T., {Oonk}, J.B. R. et al. 2015, A\&A, 580, A1
\bibitem[Morganti et al. (2017)]{morganti2017many} Morganti, R., 2017, Frontiers in Ast. and Space Sci., 4, 42
\bibitem[Mundell et al. (2009)]{mun} Mundell C. G.; Ferruit, P.; Nagar, N.; Wilson, A. S., 2009, ApJ 703, 802	
\bibitem[Nikolajuk \& Walter (2013)]{NikWal} Nikolajuk, M., \& Walter, R. 2013, A\&A 552, A75
\bibitem[Oosterloo et al. (2017)]{oosterloo2017properties} Oosterloo, T., Oonk, JB R., Morganti, R., Combes, F., Dasyra, K., 2017, A\&A, 608, A38
\bibitem[Peters et al. (2017)]{peters2017}Peters, W. et al., 2017, MNRAS, 469, 3541 
\bibitem[Ramakrishnan et al. (2019)]{10.1093/mnras/stz1244} Ramakrishnan, V., Nagar, N. M., Finlez, C. et al. 201f9, MNRAS, 487, 444
\bibitem[Raimundo et al. (2013)]{10.1093/mnras/stt327} Raimundo, S. I., Davies, R. I., Gandhi, P. et al.  2013, MNRAS, 431, 2294
\bibitem[Raimundo et al. (2017)]{raimundo2016tracing} Raimundo, S.I., Davies, R.I., Canning, R.E., et al. 2017, MNRAS, 464, 4227
\bibitem[Reynolds et al. (2000)]{2000ApJ...533..811R} {Reynolds}, Christopher S., et al. 2000, ApJ, 533, 811
\bibitem[Rosario et al. (2018)]{Rosario} Rosario D.J., et al., 2018, MNRAS 473, 5658
\bibitem[Ruffa et al. (2019)]{ruffa2019agn} Ruffa, I., Davis, T. A., Prandoni, I., et al. 2019, MNRAS, 489, 3739 
\bibitem[Rupke \& Veilleux (2013)]{rupke2013multiphase} Rupke, D. S.N., Veilleux, S., 2013, ApJ, 768, 75
\bibitem[Rush, Spinoglio, Malkan (1993)]{rush+93} Rush B.,  Malkan M.A., Spinoglio L., 1993 ApJS 89, 1
\bibitem[Sakamoto et al. (2014)]{Sakamoto_2014} Sakamoto, K., Aalto, S.,  Combes, F., Evans, A., Peck, A. 2014, ApJ, 797, 90 
\bibitem[Salak et al. (2020)]{Salak_2020} Salak, D., Nakai, N., Sorai, K., Miyamoto, Y., 2020, ApJ, 901, 151 
\bibitem[Schinnerer et al. (2000)]{Schinnerer2000}Schinnerer, E., Eckart, A., \& Tacconi, L. J. 2000, ApJ, 533, 826
\bibitem[Schmitt et al. (2003)]{2003ApJS..148..327S} {{Schmitt}, H.~R., {Donley}, J.~L., {Antonucci}, R.~R.~J.} et al. 2003, \apjs, 148, 327 
\bibitem[Sirressi et al. (2019)]{sirressi2019testing} Sirressi, M., Cicone, C., Severgnini, P. et al.  2019, MNRAS, 489, 1927
\bibitem[Slater et al. (2019)]{slater2019} {Slater, R.,} {Nagar, N. M.,} {Schnorr-M\"uller, A.}, et al. 2019, A\&A, 621, A83
\bibitem[Solomon \& Van den Bout (2005)]{solomon2005molecular}Solomon, P.M., Van de Bout, P.A., 2005, ARA\&A, 43, 677
\bibitem[Strong et al. (2004)]{strong2004molecular} Strong, M., Pedlar, A., Aalto, S. et al.  2004, MNRAS, 353, 1151
\bibitem[Tadhunter et al. (2014)]{Tadhunter2014} Tadhunter, C., Morganti, R., Rose, M., Oonk, J. B. R., Oosterloo, T., 2014, Nature, 511, 440
\bibitem[Thomas et al. (2017)]{Thomas} Thomas A.D., et al., 2017, ApJSS 232, 11
\bibitem[Tommasin et al. (2008)]{tommasin2008spitzer} Tommasin, S., Spinoglio, L., Malkan, M. A. et al. 2008, ApJ, 676, 836
\bibitem[Tommasin et al. (2010)]{tommasin2010spitzer} Tommasin, S., Spinoglio, L., Malkan, M., and Fazio, G. et al. 2010, ApJ, 709, 1257
\bibitem[Wu et al. (2009)]{2009ApJ...701..658W} {Wu, Y.}, {Charmandaris, V.}, {Huang, J.}, {Spinoglio, L.}, {Tommasin, S.} et al. 2009, ApJ, 701, 658
\bibitem[Veilleux et al. (2020)]{Veilleux2020} Veilleux, S., Maiolino, R., Bolatto, A., Aalto, S. et al. 2020, The Astronomy and Astrophysics Review, 28, 1


\end{thebibliography}
\end{document}